\documentclass[aps,twocolumn,showpacs,superscriptaddress,floatfix]{revtex4-1}

\usepackage[utf8]{inputenc}
\usepackage{braket}
\usepackage{amsmath,amssymb,graphicx}
\usepackage{qcircuit}
\usepackage{xcolor}
\usepackage{verbatim}
\usepackage[normalem]{ulem}

\DeclareMathOperator{\Tr}{Tr}

\newcommand{\zt}{\tilde{z}}

\begin{document}

\title{Many Body Thermodynamics on Quantum Computers via Partition Function Zeros}

\author{Akhil Francis}
\affiliation{Department of Physics, North Carolina State University, Raleigh, North Carolina 27695, USA}

\author{D. Zhu}
\affiliation{Joint  Quantum  Institute and  Department  of  Physics,  University  of  Maryland,  College  Park,  Maryland  20742,  USA}
\affiliation{Center for Quantum Information and Computer Science,
University of Maryland, College Park, MD 20742, USA}

\author{C. Huerta Alderete}
\affiliation{Joint  Quantum  Institute and  Department  of  Physics,  University  of  Maryland,  College  Park,  Maryland  20742,  USA}
\affiliation{Instituto Nacional de Astrof\'{i}ica, \'{O}ptica y Electr\'{o}nica, Calle Luis Enrique Erro No. 1, Sta. Ma. Tonantzintla, Pue. CP 72840, Mexico}

\author{Sonika Johri}
\affiliation{IonQ Inc., College Park, MD 20742, USA
}

\author{Xiao Xiao}
\affiliation{Department of Physics, North Carolina State University, Raleigh, North Carolina 27695, USA}

\author{J.~K.~Freericks}
\affiliation{Department of Physics, Georgetown University, 37th and O Sts. NW, Washington, DC 20057 USA}

\author{C. Monroe}
\affiliation{Joint  Quantum  Institute and  Department  of  Physics,  University  of  Maryland,  College  Park,  Maryland  20742,  USA}
\affiliation{Center for Quantum Information and Computer Science,
University of Maryland, College Park, MD 20742, USA}

\author{N. M. Linke}
\affiliation{Joint  Quantum  Institute and  Department  of  Physics,  University  of  Maryland,  College  Park,  Maryland  20742,  USA}

\author{A.~F.~Kemper}
\email{akemper@ncsu.edu}
\affiliation{Department of Physics, North Carolina State University, Raleigh, North Carolina 27695, USA}

\maketitle


{\bf 
Interacting quantum systems illustrate complex phenomena including phase transitions to novel ordered phases. The universal nature of critical phenomena reduces their description to determining only the transition temperature and the critical exponents. Numerically calculating these results for systems in new universality classes is complicated due to critical slowing down, requiring increasing resources near the critical point. An alternative approach analytically continues the calculation onto the complex plane and determines the partition function via its zeros. Here we show how to robustly perform this analysis on noisy intermediate scale trapped ion quantum computers in a scalable manner, using the XXZ model as a prototype.  We illustrate the transition from XY-like behavior to Ising-like behavior as a function of the anisotropy. While quantum computers cannot yet scale to the thermodynamic limit, our work provides a pathway to do so as hardware improves, allowing the determination of critical phenomena for systems that cannot be solved otherwise.
}

Partition functions are ubiquitous in physics. They are important in determining the thermodynamic properties of many-body systems, and in understanding their phase transitions. The partition function is real and positive;  nevertheless, its zeros can be found but only by analytically continuing the partition function to the complex plane via the introduction of complex parameters. Lee and Yang  \cite{yang_statistical_1952, yang_statistical_1952} studied the partition function zeros of Ising-like systems in the complex plane of the magnetic field $h$, and found that at the critical temperature (and in the thermodynamic limit) the loci of zeros pinch to the real axis. Alternatively, Fisher \cite{Fisher1965lectures} studied the partition function zeros by making the inverse temperature $\beta$ complex. Partition function zeros have been widely employed \cite{suzuki1971zeros,tong_lee-yang_2006} in the analysis of thermodynamic phase transitions, dynamical phase transitions \cite{heyl_dynamical_2013,brandner_experimental_2017}, and critical exponents \cite{deger_determination_2019}.
The divergence of the free energy near the phase transition is intimately connected to the location of the partition function zero closest to the real axis \cite{darboux1878memoire,hunter1980deducing}, and the critical scaling relations may be found from the density of zeros around a phase transition \cite{abe1967logarithmic}. Whenever the analytic continuation yields an analytic function in the complex plane (no poles or branch cuts), the partition function (and thus the free energy) can be reconstructed from the location of the zeros; this is typical because the partition function is a finite sum of exponentials for finite systems.

Since the zeros arise from generalizing real physical parameters to their having complex values, they were initially limited to just being useful mathematical constructs, determined either exactly for solvable systems \cite{wei_lee-yang_2012, tong_lee-yang_2006,jones1966complex,connelly2020universal} (of which there are few), or through numerical methods \cite{brandner_experimental_2017,deger2020leeyang}, which are limited by Hilbert space size in exact diagonalization
or sampling issues in Monte Carlo methods.

One notable exception was in an experimental study of the two-dimensional Ising ferromagnet, where the density of zeros was measured \cite{binek1998density}.
More recently, Liu and Wei \cite{wei_lee-yang_2012} proposed an experiment
to measure the zeros of the Ising model using the decoherence of a probe spin coupled to the Ising system; this was executed in a liquid of trimethylphosphite 
molecules using
NMR \cite{peng_experimental_2015}. While this beautifully demonstrates the technique, it is clearly not scalable as it is difficult to design molecules for every envisioned
situation. 

In this article, we employ the probe spin concept \cite{wei_lee-yang_2012} to calculate the partition function zeros on a universal quantum computer, overcoming the difficulties in numerics. 
In this manner we can handle system sizes up to the number of available qubits.
We develop a quantum circuit which evolves a thermal state \cite{wu_variational_2019,zhu_variational_2019} under a Hamiltonian consisting of an interaction with the probe spin designed to represent the action of the
complex field or temperature.
Using this,
we measure the zeros of the partition function of the XXZ model
on quantum simulators as well as trapped-ion quantum computers
as it is tuned from Ising-like to XY-like.
The locus of zeros undergoes clear {\it qualitative} changes, 
thus enabling the identification of a phase transition even on Noisy Intermediate Scale Quantum (NISQ) hardware.
With the design of the circuit being independent of a particular model,
our approach goes beyond recent studies of the Ising model \cite{krishnan_measuring_2019}.\\


\noindent {\bf Partition Function Zeros.} Our method applies to both Fisher and Lee-Yang zeros, which are zeros in the complex plane of inverse temperature $\beta$ and a complex Hamiltonian field, respectively.  First, we focus on the latter Lee-Yang case. We consider an arbitrary spin Hamiltonian ${\mathcal{H}}_s$ in the presence of an external magnetic field 
given by 
$\hat{\mathcal H}_B = h\sum_i \hat{\sigma}^z_{i}$.
As in the original work by Lee and Yang\cite{yang_statistical_1952}, the external magnetic field is complex:
$h = h_r + i h_i$. The partition function is then
\begin{align}
 \mathcal{Z}(\beta ,h) = \Tr \exp\left(- \beta {\mathcal{H}_0} - i \beta h_i \sum_i {\sigma}^z_{i}  \right),
 \end{align}
 where $\mathcal{H}_0 = \mathcal{H}_s + \mathrm{Re} (\mathcal{H}_B$). 
This expectation value is similar to that of a Loschmidt echo---the system is initially prepared in a thermal state of ${\mathcal H}_0$ and then ``time-evolved'' with respect to $\mathrm{Im}({\mathcal H}_B$).
This form suggests a direct measurement 
\begin{align}
    L(h) = \frac{1}{\mathcal{Z}_0}\Tr \exp{(-\beta {\mathcal{H}}_0 -i \beta \mathcal H_I )},
\end{align}
where $ \mathcal{Z}_{0} = \Tr e^{-\beta \mathcal{H}_0}$, and $\mathcal{H}_I=\mathrm{Im}(\mathcal{H}_B)$.
 The zeros of $L(h)$ correspond to the Lee-Yang zeros 
$\lbrace h_0 \rbrace$ of the
partition function. For a finite system of $N$ spins, we can reconstruct $\mathcal{Z}$ from its Lee-Yang zeros through
the fundamental theorem of algebra, because the partition function is a polynomial in $\zt=\exp\left(2\beta h\right)$. Hence,
\begin{align}
    \mathcal{Z}(\beta ,\zt) = \mathcal{P} \ \Pi_{j=1} ^N \left(\zt-\zt_j\right),
    \label{eq:constructZ}
\end{align}
where $\mathcal{P}$ is a numerical constant, independent of $\zt$.

As was discovered by  Wei\cite{wei_lee-yang_2012}, the quantity $L(h)$ can be measured by coupling the system
to an ancilla. 
Alternate proposals include measuring two-spin entanglement, 
but these have not yet
been realized \cite{gnatenko2017twotime,kuzmak2019probing}.
 In the simplest case, 
when $\mathcal{H}_0$ and $\mathcal{H}_I$ commute, the coupling Hamiltonian is given by
\begin{align}
	\mathcal{H}' = \frac{1}{2} \left( \sigma^z_{anc} \otimes \beta \mathcal{H}_I \right);
	\label{eq:coupling_ham}
\end{align}
for non-commuting Hamiltonians a more complex $\mathcal{H}'$ must be constructed \cite{wei_phase_2014}.
With the ancilla initialized in a superposition state $\ket{+}$ and the system in its thermal state, the
initial density matrix is
\begin{align}
    \rho(0) =(\ket{+}\bra{+}) \otimes 
    \frac{e^{-\beta \mathcal{H}_0}}{\mathcal{Z}_{0}}.
    \label{eq:rho0tot}
\end{align}
After evolution with $\mathcal{H}_I$ and tracing out the system qubits, the off-diagonal element of the ancilla density matrix
is $L(h)$.

Fisher zeros are measured using an analogous procedure; since in this case $\beta$ is complex;
the evolution is with respect to the Hamiltonian $\mathcal{H}_0$, which always commutes with itself, and hence
Eq.~(\ref{eq:coupling_ham}) always applies.\\

\begin{figure}[htpb]
\includegraphics[width=0.49\textwidth]{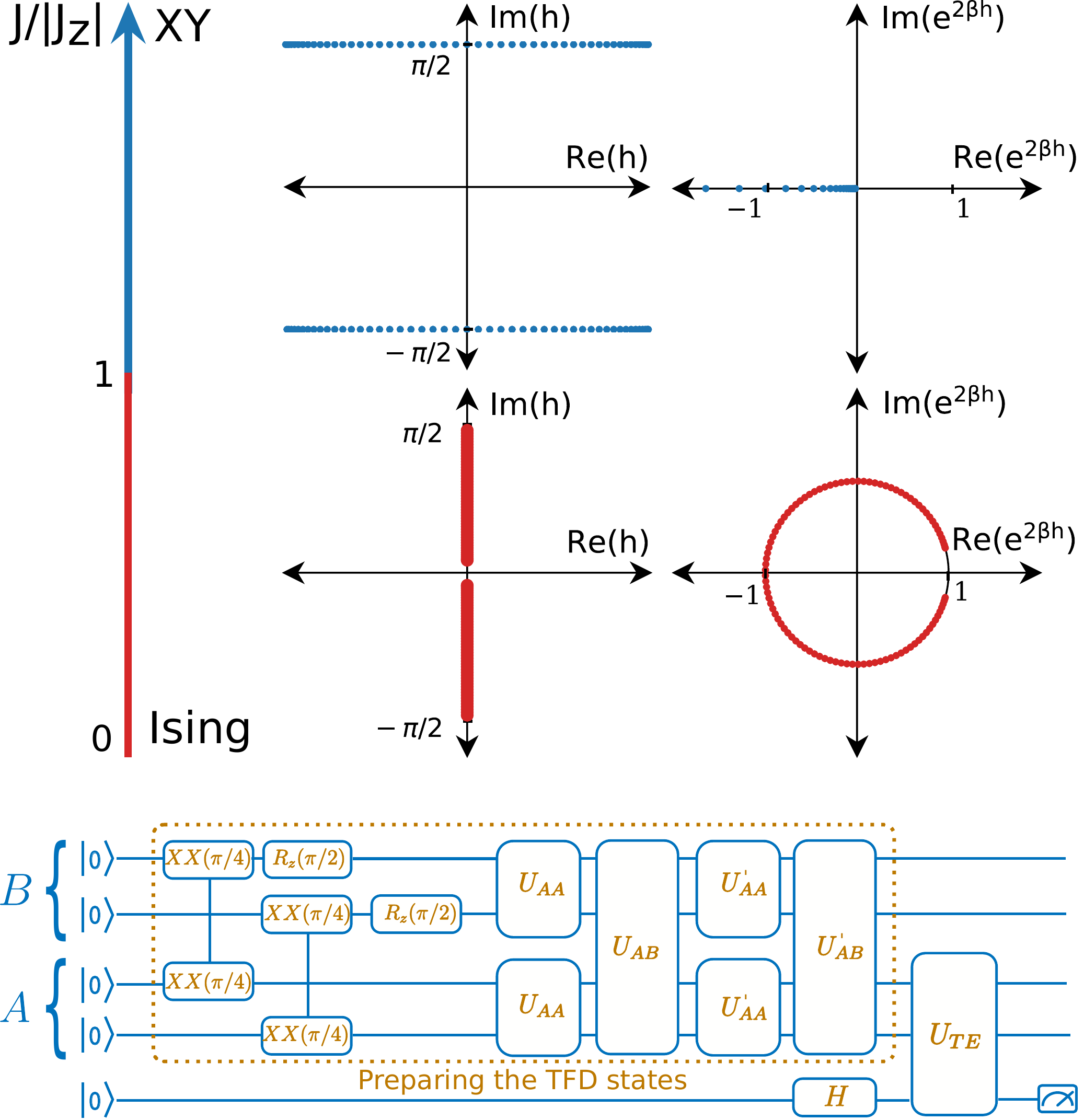}
   \caption{Top: Positions of the Lee-Yang zeros for 100-site Ising/XY models at $\beta=1$) in the complex planes of $h$ (left) and $\zt=e^{2\beta h}$ (right). For both Ising/XY models, the zeros in $h$ occur away from the real axis.  Bottom: Circuit for
   obtaining partition function zeros. The thermofield double state is prepared using a variational quantum circuit. The thermal density matrix in subsystem A is subsequently evolved under a Hamiltonian coupling it to an ancilla spin. The ancilla coherence reflects the complex partition function. The measurement operation here represents the characterization of the real/imaginary parts of the coherences (off-diagonals) of the ancilla density matrix. We achieve this through measurement in both the $x$ and $y$ basis (see Methods for details).
   }
    \label{fig:overview}
\end{figure}

\begin{figure*}[htpb]
\includegraphics[width=0.99\textwidth]{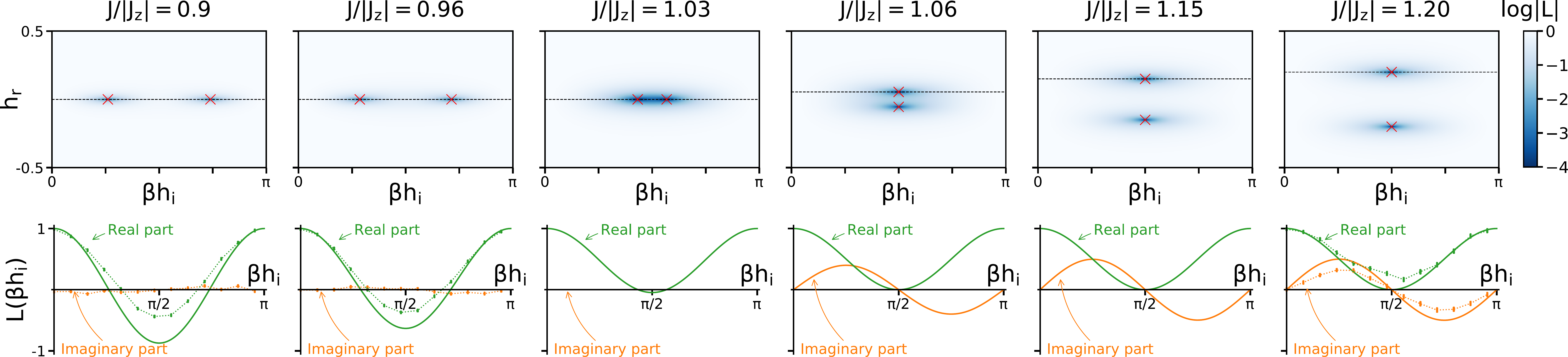}
    \caption{Phase transition from Ising to XY at $\beta=10$ as demonstrated by the nature of the Lee-Yang zeros.
        Top: False color plots of  $\log |L(h)|$ in complex $h$ space. The line in the color plot shows the $h$ values we are probing through the quantum circuit to find $L(h)$. 
    The location of the zeros are marked in the top panels with a red cross.
    Bottom: Real and imaginary parts of $L(h)$ along the cut indicated in the top row. Experimental results from the trapped ion quantum computer are shown with error bars connected with dotted lines .
        Around $J\approx |J_z|$, the nature of the zeros changes qualitatively from Ising-like to XY-like. }
        \label{fig:Figure2_experimental_data}
\end{figure*}

\noindent {\bf Model.}
We apply the above method to the one-dimensional periodic XXZ model
--- an interacting spin model that adjusts the anisotropy between spin exchange in the $x$-$y$-plane versus spin exchange along the $z$-direction --- whose Hamiltonian is given by
\begin{align}
{\mathcal H}_s = J\sum_i( \sigma^x_i \sigma^x_{i+1} +\sigma^y_i \sigma^y_{i+1} ) + J_z \sum_i \sigma^z_i \sigma^z_{i+1}.
\end{align}
We will work within the ferromagnetic Ising regime, i.e. $J_z = - |J_z|$.
The model may be tuned between an Ising-like regime ($|J_z| \gg |J|$) and an XY-like regime ($|J_z| \ll |J|$); see Fig.~\ref{fig:overview}. 
To obtain the Lee-Yang zeros, we employ a magnetic field $\mathcal{H}_B$ along the $z$ axis.

The Lee-Yang zeros of the ferromagnetic Ising model are well known
to be purely imaginary in $h$, or to lie on the complex unit circle in $\zt=\exp\left(2\beta h\right)$ \cite{lee_statistical_1952,wei_lee-yang_2012}. 
Fig.~\ref{fig:overview} shows the position of the zeros in the complex planes of $h$ and $\zt$ for
a 100-site chain.
On the other end, i.e. the XY model, the Lee-Yang zeros are qualitatively different. The zeros (in $h$) have a 
 constant imaginary component $2 \beta h_i=\left(2n + 1\right)\pi$, and their real part is given by the dispersion of the model
 after diagonalization via Jordan-Wigner transformation $h_r = - 2 J \cos(k)$ \cite{tong_lee-yang_2006-1}
 (in the quantum circuit, any finite $h_r$ must be included in the
 thermal state preparation).
 In between these limits, the zeros transition from one type to the other;
 we denote the character of the zeros as Ising-like or XY-like for the two cases, respectively.
 For zero temperature, the ground state abruptly changes from Ising-like to XY-like at $J=|J_z|$, but for finite temperatures this becomes a gradual
 change.\\

\noindent {\bf Quantum Circuit.}
The circuit is constructed in two parts. First, a thermal state corresponding to the XXZ model at finite temperature needs to be produced. For this, we prepare a thermofield double (TFD) state \cite{wu_variational_2019,zhu_variational_2019} which is a purification of the thermal Gibbs state;
it involves a doubling of the number of system qubits, half of which are then discarded to produce the thermal density matrix (see Fig.~\ref{fig:overview}). Several 
methods to prepare TFD states exist \cite{zhu_variational_2019,cottrell2019build,martyn2019product}. Here we prepare the TFD state by a variational procedure reminiscent of the Quantum Approximate Optimization Algorithm \cite{farhi2014quantum}, consisting of the application of alternating Hamiltonians within and between the subsystems of the TFD state. The parameters (angles) are optimized classically (see SI). Next, we perform evolution under the coupling Hamiltonian $\mathcal{H}_I$, which is straightforwardly implemented as controlled rotations on the ancilla. Finally, the off-diagonal elements of the ancilla are measured.
For implementation on the trapped-ion hardware, $U_{AA},U_{AB}$ and $U_{TE}$ are broken down into native $XX$ gates (see Methods and SI) and parameterized via the gate angles.\\


\noindent {\bf Implementation on Trapped-Ion Hardware.}
Fig.~\ref{fig:Figure2_experimental_data} shows the results on the $2$-site XXZ model, where we focus on the
behavior of the zeros around the phase transition at $J\approx |J_z|$;  we use $\beta=10$ and $J_z=-1$.
The figure shows the magnitude and real/imaginary
parts of the ancilla coherence $L(h)$ in the top and bottom panels, respectively.
The Lee-Yang zeros are found where both the real and imaginary parts of $L(h)$ vanish.
When $J < |J_z|$, the zeros have no real part, in
agreement with the general results for Ising-like zeros (c.f. Fig.~\ref{fig:overview}),
and for $2$ sites we expect two zeros, symmetric about $\beta h_i = \pi/2$;
these are shown in the top
panel plot of $|L(h)|$.
The bottom panel presents the real and imaginary parts of $L(h)$ as a function of $\beta h_i$ (a cut along
constant $h_r$ as indicated in the top panel), comparing the exact result and the
experimental results from the trapped-ion quantum computer.  Although the exact position of the zeros is slightly different
in the experiment, the qualitative behavior is clearly the same; $L(h)$ is entirely real, starts at unity, and changes
sign once in between $\beta h_i$ values of $0$ and $\pi/2$. As $J$ is increased towards the transition, the minimum
in the real part of $L(h)$ gets shallow.

At the phase transition, the character of the zeros changes: $\beta h_i$ becomes fixed at $\pi/2$, and the real
part $h_r$ becomes nonzero. We track $h_r$ by including it in the TFD state preparation part of the circuit, and continue to sweep $\beta h_i$ 
(indicated by horizontal lines in the top panels). This transition occurs in between $J/|J_z|$ values $1.03$ and $1.06$.
On the XY-like side of the phase transition, the real part of $L(h)$ only touches zero at $\beta h_i=\pi/2$,
and $L(h)$ acquires a non-zero imaginary part. This behavior is also captured correctly by the quantum computer;
the experimental data is shown in the rightmost panel.


\begin{figure*}[htpb]
\centering
     \includegraphics[clip=true, width=0.9\textwidth]{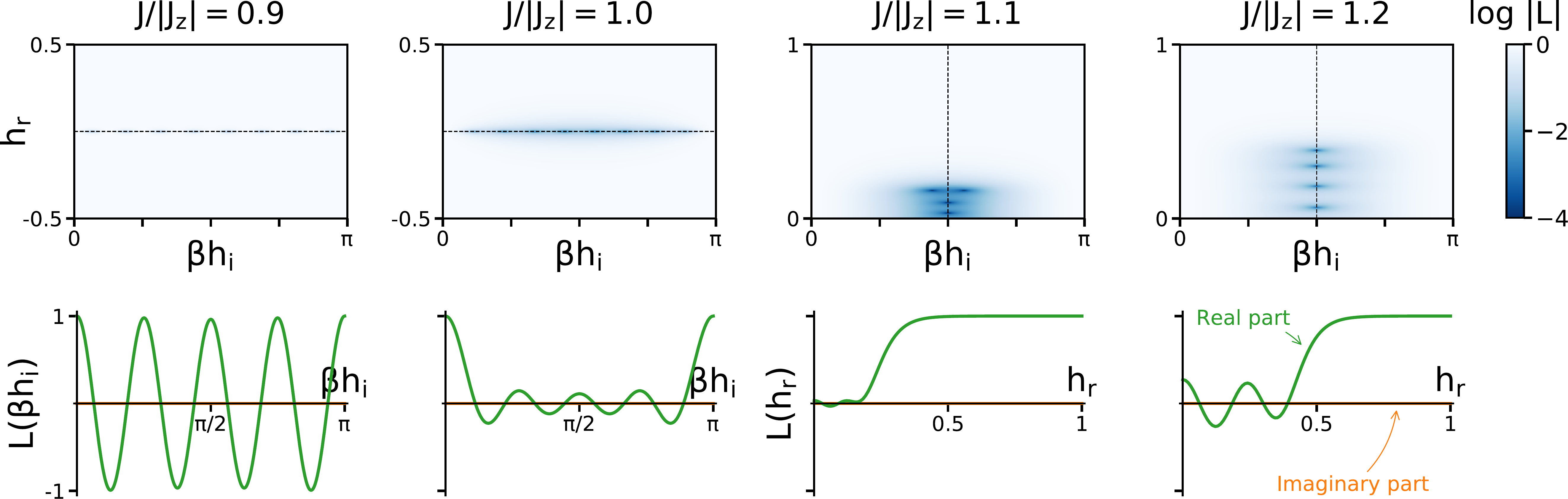}
        \caption{Lee-Yang zeros for the 8-site XXZ model. The color plots (top) show $\log |L(h)|$ in complex $h$ space. The bottom panels show cuts along the lines
        indicated in the corresponding top panel; note that for $J>|J_z|$ the cuts are vertical. 
        }
\label{Fig.3_8site_horizontal}
\end{figure*}

These data demonstrate that even with current generation NISQ hardware, a phase transition can be identified via
the qualitative character of the Lee-Yang zeros and the ancilla coherence. This is an advantage of this method;
rather than relying on a precise measurement of a quantity (such as the position of the zeros), a qualitative difference
is sufficient to distinguish the Ising-like from the XY-like regime of the model. Understanding the effect of the noise in the quantum computer on the results can further help to predict the accuracy of the locations of the zeros as the system size grows larger (see SI).

Although here we have chosen to continue to sweep along $\beta h_i$ and fix $h_r$, to avoid having to know
an exact value of $h_r$, a sweep along constant $h_i$ could be performed. More generally, if nothing is known
about the position of the zeros, a full scan of complex $h$ is possible - although time consuming, the effort does
not increase with system size.

Our approach scales readily to larger systems.  In Fig.~\ref{Fig.3_8site_horizontal} we show the results when the method is applied to an 8-site system. Due to hardware limitations
(this calculation would require 17 qubits and a similarly larger number of gates for the TFD state preparation),
only simulator data is available.

The number of zeros is now larger, and they exhibit a more complex pattern, in particular around the phase transition.  However, the overall qualitative difference between the two states
remains clear---the zeros obtain a real part and shift to lie purely along the line $\beta h_i =\pi/2$. \\

\noindent {\bf Fisher Zeros.}
%
\begin{figure}[b]
\includegraphics[width=0.98\columnwidth]{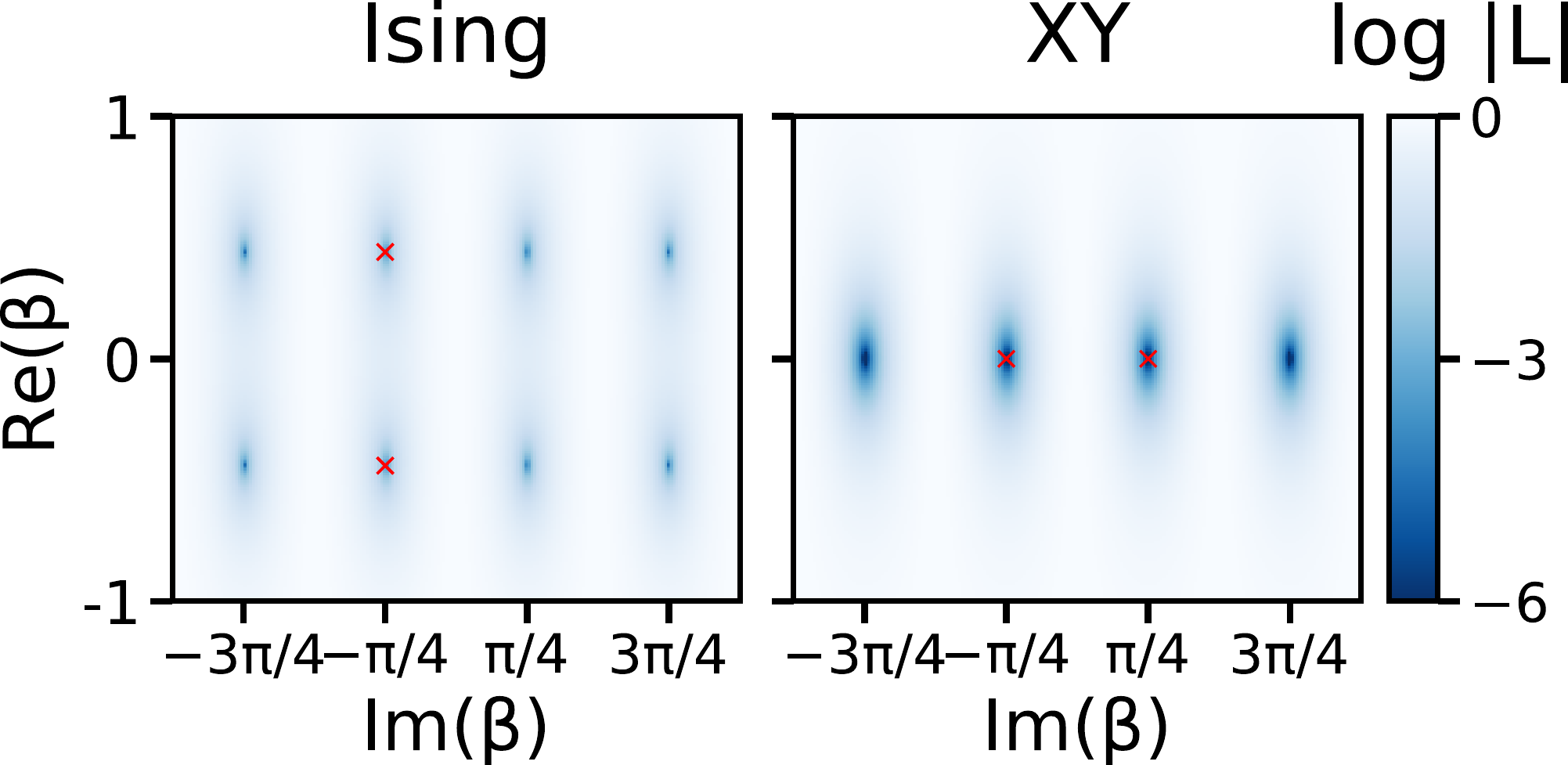}

	\caption{Fisher zeros for the two limiting cases of a 4-site XXZ model.}
	\label{fig:Fisher}
\end{figure}
The approach above can be equally applied to finding zeros in complex $\beta=T^{-1}$. In this case, the interaction Hamiltonian for the ``time evolution'' portion of
the circuit is simply 
	$\mathcal{H}' = \frac{1}{2} \left( \sigma^z_{anc} \otimes \mathcal{H}_0 \right)$,
and this amounts to the application of a controlled unitary. When $\mathcal{H}_0$ is simple, this may be implemented without approximation, but for larger or more complex
systems a Trotter decomposition of $\exp(-i \mathcal{H}_0 \beta_i)$ may be necessary.

For the two limiting cases under consideration---the XY model and the Ising model--- the Fisher zeros also show a qualitative transition. Fig.~\ref{fig:Fisher} shows the location of the zeros for the two limits.
The Ising model Fisher zeros lie parallel to the real axis\cite{jones1966complex} (similar to the XY Lee-Yang zeros), and the XY model Fisher zeros lie
directly on the imaginary axis.  In between, the features are more complex than the Lee-Yang zeros are,
and depend heavily on the choice of boundary conditions (see SI); here we have chosen the boundary conditions that make the system
amenable to a Jordan-Wigner transformation.\\


\noindent {\bf Discussion.} 
Our results show that the zeros of the partition function may be obtained by time evolution of a suitably
prepared thermal state under a Hamiltonian corresponding to either an external field (in the Lee-Yang case) or
under the Hamiltonian (in the Fisher case).  At this point, a question arises: aside from identifying phases,
what else can be done with this information?

\begin{figure}[htpb]
    \centering
     \includegraphics[width=0.4\textwidth]{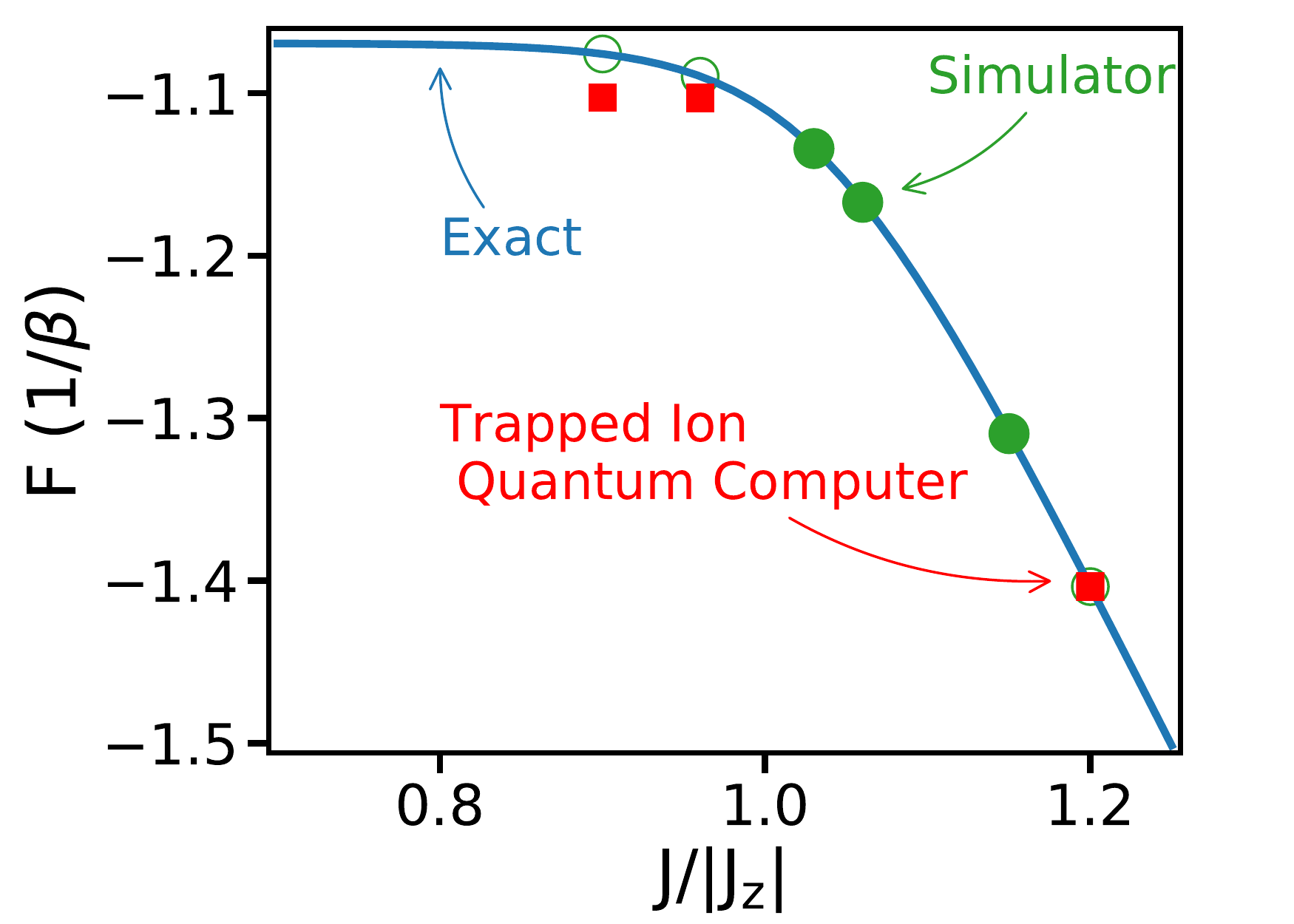}
    \caption{Free energy of the 2 site XXZ model at $\beta=10$ reconstructed from the Lee-Yang zeros. Green circles indicate $F$ as reconstructed from the quantum simulator; open circles indicates values for which experimental
    data is available and solid ones for simulated data only. Red squares indicate $F$ constructed using the experimental data.
    We have used the post selected data (Method 1, see SI) to compute the free energy.}
    \label{fig:2sitexxz_Free_energy}
\end{figure}

One obvious path is to reconstruct the free energy, given that the polynomial expansion of the partition function is known. Although
in principle this involves the evaluation of an infinite sum, when 
a closed
form can be obtained the full thermodynamics are determined.
Here this is accomplished by considering the polynomial in $\zt\equiv\exp(2\beta h_i)$ instead of $h_i$ (this is because the energy spacing of the initial spin model is uniform).
From the results shown in Fig.~\ref{fig:Figure2_experimental_data}, we can extract the set of Lee-Yang zeros $\lbrace h_0 \rbrace$,
or equivalently $\lbrace \zt_0 \rbrace$
once the prefactor $\mathcal{P}$ in the polynomial expansion 
is determined (see SI).
From the partition function, we compute the free energy $F=-(1/\beta)\ln\mathcal{Z}(\beta,h=0)$, shown in  
Fig.~\ref{fig:Figure2_experimental_data}.
In the experimental data $L(h)$ is never precisely zero; instead, we find the value of $h_i$ corresponding to the 
smallest value of $|L(h)|$ by linear interpolation between
the data points
($h_r$ is assumed to be known from the input to the TFD).
The reconstructed free energy is shown in Fig.~\ref{fig:2sitexxz_Free_energy}. The results from the quantum simulator
reproduce the correct values; the free energy obtained from the experimental results exhibits some deviations.

Beyond exactly reconstructing the partition function, the zeros also yield information regarding the thermodynamic properties
near the phase transition.  The partition function zeros lead to divergences in the free energy; moreover, in the limit as
$N\rightarrow\infty$ the zeros form a branch cut ending in an edge singularity. It is known from complex analysis that the
knowledge of a function around its branch cuts is sufficient to determine the entire function.  Furthermore, even if only
the first zero (or edge singularity) is known, the temperature and field dependence of thermodynamic functions is dominated
by its position \cite{darboux1878memoire,hunter1980deducing}: in the expansion of a complex function near a 
singularity, the terms after some order $n$ are determined by the properties of the singularity.
And finally, Abe \cite{abe1967logarithmic} showed that the dependence of density of zeros on the system size (i.e. finite size scaling)
can be used to determine critical exponents of a phase transition. The limiting density of zeros may also be used to characterise
the phase transition \cite{grossmann_temperature_1967}.
Hence, calculations on a quantum computer, focusing on zeros near the transition temperature, can efficiently determine much of the critical phenomena at the transition, perhaps easier than any other methodology.
In this work, we have outlined how partition function zeros may be obtained, and demonstrated that this is
feasible even on NISQ hardware. Thus, as improvements to quantum computers enable ever growing complex calculations, partition function zeros
can play a broadly applicable role in the simulation of physics at zero and finite temperatures.  They may be evaluated with
relative ease given a model Hamiltonian or external field, and yield a wealth of information regarding the thermodynamics
of the system under study.  
\section*{Acknowledgments}
We would like to acknowledge V. Skokov for
enlightening discussions regarding partition function zeros and K. Klymko for comments.
This work was supported by the Department of Energy, Office of Basic Energy Sciences, Division of Materials Sciences and Engineering under Grant No. DE-SC0019469. 
J.K.F. was also supported by the McDevitt bequest at Georgetown.
C.H.A. acknowledges financial support from CONACYT doctoral grant No. 455378.
N.M.L. acknowledges financial support from the NSF grant no. PHY-1430094 to the
PFC@JQI. We acknowledge the use of Qiskit for simulations\cite{Qiskit}, and acknowledge the use of IBMQ via the IBM Q Hub at NC State for this work.  The views expressed are those of the authors and do not reflect the official policy or position of the IBM Q Hub at NC State, IBM or the IBM Q team.

\section*{Author Contributions}
A.F.K. conceptualized the project. 
A.F., A.F.K, and S.J. designed and optimized the circuit, and executed simulations. D.Z., C.H.A. and N.M.L. performed the experimental trapped-ion measurements.  
A.F., S.J., and D.Z. analyzed simulator and experimental data.
X.X. provided valuable context to the discussion.
All authors discussed the results and contributed to the development of the manuscript.

\section*{Methods}
We use a re-configurable digital quantum computer for this study. The system is made of a chain of  ${}^{171}\textrm{Yb}^+$ ions trapped with radio frequency electric field \cite{debnath2016demonstration}. The pair of states in the hyperfine-split ${}^{2}\textrm{S}_{1/2}$ ground level of each ion, connected by a magnetic field insensitive 12.642821 GHz transition, is used as a physical qubit. Optical pumping are used to initialize qubits to $\ket{0}$. The read out on the other hand is implemented through state-dependent fluorescence detection \cite{Olmschenk07}.
The complete set of quantum gates is realized with a pair of Raman beams derived from a single 355-nm mode-locked laser. Our native single-qubit gates are rotations along arbitrary axis of the Bloch sphere for arbitrary angles. This is achieved by driving resonant Rabi-transitions between the two qubit states. Our native Two-qubit gates are XX (Ising) gate implemented using the phonon-mediated Molmer-Sorensen interaction \cite{Molmer99,Solano99}. To reach optimal performance, our scheme utilizes multiple phonon modes, which are disentangled from the qubits at the end of an two-qubit gate operations via an amplitude modulation scheme\cite{choi2014optimal}.Our single- and two-qubit gate fidelities are typically around $99.5(2)\%$ and $98-99\%$, respectively. The residual entanglement between the qubit states and the phonon state is the major factor limiting the fidelity of each two-qubit gates. However, for circuits with more than 10 two-qubit gates, mode-heating of phonon modes also plays a significant role. 

Our native XX gate is defined as $XX(\theta)=e^{-i \sigma_x^{(j)} \sigma_x^{(k)}\theta }$, where $j,k$ correspond to the two ions the gate is applied on. We convert this XX gate to YY and ZZ gates, defined as $YY(\theta)=e^{-i \sigma_y^{(j)} \sigma_y^{(k)}\theta }$ and $ZZ(\theta)=e^{-i \sigma_z^{(j)} \sigma_z^{(k)}\theta }$, respectively, by rotating the interaction axis through single qubit rotations. Here, $\sigma_\alpha^{(k)}$ is the
$\alpha$-th Pauli matrix applied to the $k$-th qubit.

The readout operations are simultaneously performed on all the qubits. The measurements are, by default, in the computational(z) basis. We append the circuit with an additional $R_y(-\pi/2)$($R_x(\pi/2)$) rotation to perform measurements in $x$($y$) basis.

\bibliography{main}

\begin{thebibliography}{34}%
\makeatletter
\providecommand \@ifxundefined [1]{%
 \@ifx{#1\undefined}
}%
\providecommand \@ifnum [1]{%
 \ifnum #1\expandafter \@firstoftwo
 \else \expandafter \@secondoftwo
 \fi
}%
\providecommand \@ifx [1]{%
 \ifx #1\expandafter \@firstoftwo
 \else \expandafter \@secondoftwo
 \fi
}%
\providecommand \natexlab [1]{#1}%
\providecommand \enquote  [1]{``#1''}%
\providecommand \bibnamefont  [1]{#1}%
\providecommand \bibfnamefont [1]{#1}%
\providecommand \citenamefont [1]{#1}%
\providecommand \href@noop [0]{\@secondoftwo}%
\providecommand \href [0]{\begingroup \@sanitize@url \@href}%
\providecommand \@href[1]{\@@startlink{#1}\@@href}%
\providecommand \@@href[1]{\endgroup#1\@@endlink}%
\providecommand \@sanitize@url [0]{\catcode `\\12\catcode `\$12\catcode
  `\&12\catcode `\#12\catcode `\^12\catcode `\_12\catcode `\%12\relax}%
\providecommand \@@startlink[1]{}%
\providecommand \@@endlink[0]{}%
\providecommand \url  [0]{\begingroup\@sanitize@url \@url }%
\providecommand \@url [1]{\endgroup\@href {#1}{\urlprefix }}%
\providecommand \urlprefix  [0]{URL }%
\providecommand \Eprint [0]{\href }%
\providecommand \doibase [0]{http://dx.doi.org/}%
\providecommand \selectlanguage [0]{\@gobble}%
\providecommand \bibinfo  [0]{\@secondoftwo}%
\providecommand \bibfield  [0]{\@secondoftwo}%
\providecommand \translation [1]{[#1]}%
\providecommand \BibitemOpen [0]{}%
\providecommand \bibitemStop [0]{}%
\providecommand \bibitemNoStop [0]{.\EOS\space}%
\providecommand \EOS [0]{\spacefactor3000\relax}%
\providecommand \BibitemShut  [1]{\csname bibitem#1\endcsname}%
\let\auto@bib@innerbib\@empty
\bibitem [{\citenamefont {Yang}\ and\ \citenamefont
  {Lee}(1952)}]{yang_statistical_1952}%
  \BibitemOpen
  \bibfield  {author} {\bibinfo {author} {\bibfnamefont {C.~N.}\ \bibnamefont
  {Yang}}\ and\ \bibinfo {author} {\bibfnamefont {T.~D.}\ \bibnamefont {Lee}},\
  }\href {\doibase 10.1103/PhysRev.87.404} {\bibfield  {journal} {\bibinfo
  {journal} {Physical Review}\ }\textbf {\bibinfo {volume} {87}},\ \bibinfo
  {pages} {404} (\bibinfo {year} {1952})}\BibitemShut {NoStop}%
\bibitem [{\citenamefont {Fisher}(1965)}]{Fisher1965lectures}%
  \BibitemOpen
  \bibfield  {author} {\bibinfo {author} {\bibfnamefont {M.~E.}\ \bibnamefont
  {Fisher}},\ }\href@noop {} {\bibfield  {journal} {\bibinfo  {journal} {Vol.
  VU C (University of Colorado Press 1965)}\ } (\bibinfo {year}
  {1965})}\BibitemShut {NoStop}%
\bibitem [{\citenamefont {Suzuki}\ and\ \citenamefont
  {Fisher}(1971)}]{suzuki1971zeros}%
  \BibitemOpen
  \bibfield  {author} {\bibinfo {author} {\bibfnamefont {M.}~\bibnamefont
  {Suzuki}}\ and\ \bibinfo {author} {\bibfnamefont {M.~E.}\ \bibnamefont
  {Fisher}},\ }\href@noop {} {\bibfield  {journal} {\bibinfo  {journal}
  {Journal of Mathematical Physics}\ }\textbf {\bibinfo {volume} {12}},\
  \bibinfo {pages} {235} (\bibinfo {year} {1971})}\BibitemShut {NoStop}%
\bibitem [{\citenamefont {Tong}\ and\ \citenamefont
  {Liu}(2006{\natexlab{a}})}]{tong_lee-yang_2006}%
  \BibitemOpen
  \bibfield  {author} {\bibinfo {author} {\bibfnamefont {P.}~\bibnamefont
  {Tong}}\ and\ \bibinfo {author} {\bibfnamefont {X.}~\bibnamefont {Liu}},\
  }\href {\doibase 10.1103/PhysRevLett.97.017201} {\bibfield  {journal}
  {\bibinfo  {journal} {Physical Review Letters}\ }\textbf {\bibinfo {volume}
  {97}},\ \bibinfo {pages} {017201} (\bibinfo {year}
  {2006}{\natexlab{a}})}\BibitemShut {NoStop}%
\bibitem [{\citenamefont {Heyl}\ \emph {et~al.}(2013)\citenamefont {Heyl},
  \citenamefont {Polkovnikov},\ and\ \citenamefont
  {Kehrein}}]{heyl_dynamical_2013}%
  \BibitemOpen
  \bibfield  {author} {\bibinfo {author} {\bibfnamefont {M.}~\bibnamefont
  {Heyl}}, \bibinfo {author} {\bibfnamefont {A.}~\bibnamefont {Polkovnikov}}, \
  and\ \bibinfo {author} {\bibfnamefont {S.}~\bibnamefont {Kehrein}},\ }\href
  {\doibase 10.1103/PhysRevLett.110.135704} {\bibfield  {journal} {\bibinfo
  {journal} {Physical Review Letters}\ }\textbf {\bibinfo {volume} {110}},\
  \bibinfo {pages} {135704} (\bibinfo {year} {2013})}\BibitemShut {NoStop}%
\bibitem [{\citenamefont {Brandner}\ \emph {et~al.}(2017)\citenamefont
  {Brandner}, \citenamefont {Maisi}, \citenamefont {Pekola}, \citenamefont
  {Garrahan},\ and\ \citenamefont {Flindt}}]{brandner_experimental_2017}%
  \BibitemOpen
  \bibfield  {author} {\bibinfo {author} {\bibfnamefont {K.}~\bibnamefont
  {Brandner}}, \bibinfo {author} {\bibfnamefont {V.~F.}\ \bibnamefont {Maisi}},
  \bibinfo {author} {\bibfnamefont {J.~P.}\ \bibnamefont {Pekola}}, \bibinfo
  {author} {\bibfnamefont {J.~P.}\ \bibnamefont {Garrahan}}, \ and\ \bibinfo
  {author} {\bibfnamefont {C.}~\bibnamefont {Flindt}},\ }\href {\doibase
  10.1103/PhysRevLett.118.180601} {\bibfield  {journal} {\bibinfo  {journal}
  {Physical Review Letters}\ }\textbf {\bibinfo {volume} {118}},\ \bibinfo
  {pages} {180601} (\bibinfo {year} {2017})}\BibitemShut {NoStop}%
\bibitem [{\citenamefont {Deger}\ and\ \citenamefont
  {Flindt}(2019)}]{deger_determination_2019}%
  \BibitemOpen
  \bibfield  {author} {\bibinfo {author} {\bibfnamefont {A.}~\bibnamefont
  {Deger}}\ and\ \bibinfo {author} {\bibfnamefont {C.}~\bibnamefont {Flindt}},\
  }\href {\doibase 10.1103/PhysRevResearch.1.023004} {\bibfield  {journal}
  {\bibinfo  {journal} {Physical Review Research}\ }\textbf {\bibinfo {volume}
  {1}} (\bibinfo {year} {2019}),\ 10.1103/PhysRevResearch.1.023004}\BibitemShut
  {NoStop}%
\bibitem [{\citenamefont {Darboux}(1878)}]{darboux1878memoire}%
  \BibitemOpen
  \bibfield  {author} {\bibinfo {author} {\bibfnamefont {G.}~\bibnamefont
  {Darboux}},\ }\href@noop {} {\bibfield  {journal} {\bibinfo  {journal}
  {Journal de Math{\'e}matiques pures et appliqu{\'e}es}\ ,\ \bibinfo {pages}
  {5}} (\bibinfo {year} {1878})}\BibitemShut {NoStop}%
\bibitem [{\citenamefont {Hunter}\ and\ \citenamefont
  {Guerrieri}(1980)}]{hunter1980deducing}%
  \BibitemOpen
  \bibfield  {author} {\bibinfo {author} {\bibfnamefont {C.}~\bibnamefont
  {Hunter}}\ and\ \bibinfo {author} {\bibfnamefont {B.}~\bibnamefont
  {Guerrieri}},\ }\href@noop {} {\bibfield  {journal} {\bibinfo  {journal}
  {SIAM Journal on Applied Mathematics}\ }\textbf {\bibinfo {volume} {39}},\
  \bibinfo {pages} {248} (\bibinfo {year} {1980})}\BibitemShut {NoStop}%
\bibitem [{\citenamefont {Abe}(1967)}]{abe1967logarithmic}%
  \BibitemOpen
  \bibfield  {author} {\bibinfo {author} {\bibfnamefont {R.}~\bibnamefont
  {Abe}},\ }\href@noop {} {\bibfield  {journal} {\bibinfo  {journal} {Progress
  of Theoretical Physics}\ }\textbf {\bibinfo {volume} {37}},\ \bibinfo {pages}
  {1070} (\bibinfo {year} {1967})}\BibitemShut {NoStop}%
\bibitem [{\citenamefont {Wei}\ and\ \citenamefont
  {Liu}(2012)}]{wei_lee-yang_2012}%
  \BibitemOpen
  \bibfield  {author} {\bibinfo {author} {\bibfnamefont {B.-B.}\ \bibnamefont
  {Wei}}\ and\ \bibinfo {author} {\bibfnamefont {R.-B.}\ \bibnamefont {Liu}},\
  }\href {\doibase 10.1103/PhysRevLett.109.185701} {\bibfield  {journal}
  {\bibinfo  {journal} {Physical Review Letters}\ }\textbf {\bibinfo {volume}
  {109}},\ \bibinfo {pages} {185701} (\bibinfo {year} {2012})}\BibitemShut
  {NoStop}%
\bibitem [{\citenamefont {Jones}(1966)}]{jones1966complex}%
  \BibitemOpen
  \bibfield  {author} {\bibinfo {author} {\bibfnamefont {G.~L.}\ \bibnamefont
  {Jones}},\ }\href@noop {} {\bibfield  {journal} {\bibinfo  {journal} {Journal
  of Mathematical Physics}\ }\textbf {\bibinfo {volume} {7}},\ \bibinfo {pages}
  {2000} (\bibinfo {year} {1966})}\BibitemShut {NoStop}%
\bibitem [{\citenamefont {Connelly}\ \emph {et~al.}(2020)\citenamefont
  {Connelly}, \citenamefont {Johnson}, \citenamefont {Rennecke},\ and\
  \citenamefont {Skokov}}]{connelly2020universal}%
  \BibitemOpen
  \bibfield  {author} {\bibinfo {author} {\bibfnamefont {A.}~\bibnamefont
  {Connelly}}, \bibinfo {author} {\bibfnamefont {G.}~\bibnamefont {Johnson}},
  \bibinfo {author} {\bibfnamefont {F.}~\bibnamefont {Rennecke}}, \ and\
  \bibinfo {author} {\bibfnamefont {V.}~\bibnamefont {Skokov}},\ }\href@noop {}
  {\bibfield  {journal} {\bibinfo  {journal} {arXiv preprint arXiv:2006.12541}\
  } (\bibinfo {year} {2020})}\BibitemShut {NoStop}%
\bibitem [{\citenamefont {Deger}\ and\ \citenamefont
  {Flindt}(2020)}]{deger2020leeyang}%
  \BibitemOpen
  \bibfield  {author} {\bibinfo {author} {\bibfnamefont {A.}~\bibnamefont
  {Deger}}\ and\ \bibinfo {author} {\bibfnamefont {C.}~\bibnamefont {Flindt}},\
  }\href {\doibase 10.1103/PhysRevResearch.2.033009} {\bibfield  {journal}
  {\bibinfo  {journal} {Phys. Rev. Research}\ }\textbf {\bibinfo {volume}
  {2}},\ \bibinfo {pages} {033009} (\bibinfo {year} {2020})}\BibitemShut
  {NoStop}%
\bibitem [{\citenamefont {Binek}(1998)}]{binek1998density}%
  \BibitemOpen
  \bibfield  {author} {\bibinfo {author} {\bibfnamefont {C.}~\bibnamefont
  {Binek}},\ }\href {\doibase 10.1103/PhysRevLett.81.5644} {\bibfield
  {journal} {\bibinfo  {journal} {Phys. Rev. Lett.}\ }\textbf {\bibinfo
  {volume} {81}},\ \bibinfo {pages} {5644} (\bibinfo {year}
  {1998})}\BibitemShut {NoStop}%
\bibitem [{\citenamefont {Peng}\ \emph {et~al.}(2015)\citenamefont {Peng},
  \citenamefont {Zhou}, \citenamefont {Wei}, \citenamefont {Cui}, \citenamefont
  {Du},\ and\ \citenamefont {Liu}}]{peng_experimental_2015}%
  \BibitemOpen
  \bibfield  {author} {\bibinfo {author} {\bibfnamefont {X.}~\bibnamefont
  {Peng}}, \bibinfo {author} {\bibfnamefont {H.}~\bibnamefont {Zhou}}, \bibinfo
  {author} {\bibfnamefont {B.-B.}\ \bibnamefont {Wei}}, \bibinfo {author}
  {\bibfnamefont {J.}~\bibnamefont {Cui}}, \bibinfo {author} {\bibfnamefont
  {J.}~\bibnamefont {Du}}, \ and\ \bibinfo {author} {\bibfnamefont {R.-B.}\
  \bibnamefont {Liu}},\ }\href {\doibase 10.1103/PhysRevLett.114.010601}
  {\bibfield  {journal} {\bibinfo  {journal} {Physical Review Letters}\
  }\textbf {\bibinfo {volume} {114}} (\bibinfo {year} {2015}),\
  10.1103/PhysRevLett.114.010601}\BibitemShut {NoStop}%
\bibitem [{\citenamefont {Wu}\ and\ \citenamefont
  {Hsieh}(2019)}]{wu_variational_2019}%
  \BibitemOpen
  \bibfield  {author} {\bibinfo {author} {\bibfnamefont {J.}~\bibnamefont
  {Wu}}\ and\ \bibinfo {author} {\bibfnamefont {T.~H.}\ \bibnamefont {Hsieh}},\
  }\href {\doibase 10.1103/PhysRevLett.123.220502} {\bibfield  {journal}
  {\bibinfo  {journal} {Physical Review Letters}\ }\textbf {\bibinfo {volume}
  {123}},\ \bibinfo {pages} {220502} (\bibinfo {year} {2019})}\BibitemShut
  {NoStop}%
\bibitem [{\citenamefont {Zhu}\ \emph {et~al.}(2019)\citenamefont {Zhu},
  \citenamefont {Johri}, \citenamefont {Linke}, \citenamefont {Landsman},
  \citenamefont {Nguyen}, \citenamefont {Alderete}, \citenamefont {Matsuura},
  \citenamefont {Hsieh},\ and\ \citenamefont {Monroe}}]{zhu_variational_2019}%
  \BibitemOpen
  \bibfield  {author} {\bibinfo {author} {\bibfnamefont {D.}~\bibnamefont
  {Zhu}}, \bibinfo {author} {\bibfnamefont {S.}~\bibnamefont {Johri}}, \bibinfo
  {author} {\bibfnamefont {N.~M.}\ \bibnamefont {Linke}}, \bibinfo {author}
  {\bibfnamefont {K.~A.}\ \bibnamefont {Landsman}}, \bibinfo {author}
  {\bibfnamefont {N.~H.}\ \bibnamefont {Nguyen}}, \bibinfo {author}
  {\bibfnamefont {C.~H.}\ \bibnamefont {Alderete}}, \bibinfo {author}
  {\bibfnamefont {A.~Y.}\ \bibnamefont {Matsuura}}, \bibinfo {author}
  {\bibfnamefont {T.~H.}\ \bibnamefont {Hsieh}}, \ and\ \bibinfo {author}
  {\bibfnamefont {C.}~\bibnamefont {Monroe}},\ }\href
  {http://arxiv.org/abs/1906.02699} {\bibfield  {journal} {\bibinfo  {journal}
  {arXiv:1906.02699 [cond-mat, physics:hep-th, physics:quant-ph]}\ } (\bibinfo
  {year} {2019})},\ \bibinfo {note} {arXiv: 1906.02699}\BibitemShut {NoStop}%
\bibitem [{\citenamefont {Krishnan}\ \emph {et~al.}(2019)\citenamefont
  {Krishnan}, \citenamefont {Schmitt}, \citenamefont {Moessner},\ and\
  \citenamefont {Heyl}}]{krishnan_measuring_2019}%
  \BibitemOpen
  \bibfield  {author} {\bibinfo {author} {\bibfnamefont {A.}~\bibnamefont
  {Krishnan}}, \bibinfo {author} {\bibfnamefont {M.}~\bibnamefont {Schmitt}},
  \bibinfo {author} {\bibfnamefont {R.}~\bibnamefont {Moessner}}, \ and\
  \bibinfo {author} {\bibfnamefont {M.}~\bibnamefont {Heyl}},\ }\href {\doibase
  10.1103/PhysRevA.100.022125} {\bibfield  {journal} {\bibinfo  {journal}
  {Physical Review A}\ }\textbf {\bibinfo {volume} {100}},\ \bibinfo {pages}
  {022125} (\bibinfo {year} {2019})}\BibitemShut {NoStop}%
\bibitem [{\citenamefont {Gnatenko}\ \emph {et~al.}(2017)\citenamefont
  {Gnatenko}, \citenamefont {Kargol},\ and\ \citenamefont
  {Tkachuk}}]{gnatenko2017twotime}%
  \BibitemOpen
  \bibfield  {author} {\bibinfo {author} {\bibfnamefont {K.~P.}\ \bibnamefont
  {Gnatenko}}, \bibinfo {author} {\bibfnamefont {A.}~\bibnamefont {Kargol}}, \
  and\ \bibinfo {author} {\bibfnamefont {V.~M.}\ \bibnamefont {Tkachuk}},\
  }\href {\doibase 10.1103/PhysRevE.96.032116} {\bibfield  {journal} {\bibinfo
  {journal} {Phys. Rev. E}\ }\textbf {\bibinfo {volume} {96}},\ \bibinfo
  {pages} {032116} (\bibinfo {year} {2017})}\BibitemShut {NoStop}%
\bibitem [{\citenamefont {Kuzmak}\ and\ \citenamefont
  {Tkachuk}(2019)}]{kuzmak2019probing}%
  \BibitemOpen
  \bibfield  {author} {\bibinfo {author} {\bibfnamefont {A.}~\bibnamefont
  {Kuzmak}}\ and\ \bibinfo {author} {\bibfnamefont {V.}~\bibnamefont
  {Tkachuk}},\ }\href@noop {} {\bibfield  {journal} {\bibinfo  {journal}
  {Journal of Physics B: Atomic, Molecular and Optical Physics}\ }\textbf
  {\bibinfo {volume} {52}},\ \bibinfo {pages} {205501} (\bibinfo {year}
  {2019})}\BibitemShut {NoStop}%
\bibitem [{\citenamefont {Wei}\ \emph {et~al.}(2014)\citenamefont {Wei},
  \citenamefont {Chen}, \citenamefont {Po},\ and\ \citenamefont
  {Liu}}]{wei_phase_2014}%
  \BibitemOpen
  \bibfield  {author} {\bibinfo {author} {\bibfnamefont {B.-B.}\ \bibnamefont
  {Wei}}, \bibinfo {author} {\bibfnamefont {S.-W.}\ \bibnamefont {Chen}},
  \bibinfo {author} {\bibfnamefont {H.-C.}\ \bibnamefont {Po}}, \ and\ \bibinfo
  {author} {\bibfnamefont {R.-B.}\ \bibnamefont {Liu}},\ }\href {\doibase
  10.1038/srep05202} {\bibfield  {journal} {\bibinfo  {journal} {Scientific
  Reports}\ }\textbf {\bibinfo {volume} {4}},\ \bibinfo {pages} {1} (\bibinfo
  {year} {2014})}\BibitemShut {NoStop}%
\bibitem [{\citenamefont {Lee}\ and\ \citenamefont
  {Yang}(1952)}]{lee_statistical_1952}%
  \BibitemOpen
  \bibfield  {author} {\bibinfo {author} {\bibfnamefont {T.~D.}\ \bibnamefont
  {Lee}}\ and\ \bibinfo {author} {\bibfnamefont {C.~N.}\ \bibnamefont {Yang}},\
  }\href {\doibase 10.1103/PhysRev.87.410} {\bibfield  {journal} {\bibinfo
  {journal} {Physical Review}\ }\textbf {\bibinfo {volume} {87}},\ \bibinfo
  {pages} {410} (\bibinfo {year} {1952})}\BibitemShut {NoStop}%
\bibitem [{\citenamefont {Tong}\ and\ \citenamefont
  {Liu}(2006{\natexlab{b}})}]{tong_lee-yang_2006-1}%
  \BibitemOpen
  \bibfield  {author} {\bibinfo {author} {\bibfnamefont {P.}~\bibnamefont
  {Tong}}\ and\ \bibinfo {author} {\bibfnamefont {X.}~\bibnamefont {Liu}},\
  }\href {\doibase 10.1103/PhysRevLett.97.017201} {\bibfield  {journal}
  {\bibinfo  {journal} {Physical Review Letters}\ }\textbf {\bibinfo {volume}
  {97}},\ \bibinfo {pages} {017201} (\bibinfo {year}
  {2006}{\natexlab{b}})}\BibitemShut {NoStop}%
\bibitem [{\citenamefont {Cottrell}\ \emph {et~al.}(2019)\citenamefont
  {Cottrell}, \citenamefont {Freivogel}, \citenamefont {Hofman},\ and\
  \citenamefont {Lokhande}}]{cottrell2019build}%
  \BibitemOpen
  \bibfield  {author} {\bibinfo {author} {\bibfnamefont {W.}~\bibnamefont
  {Cottrell}}, \bibinfo {author} {\bibfnamefont {B.}~\bibnamefont {Freivogel}},
  \bibinfo {author} {\bibfnamefont {D.~M.}\ \bibnamefont {Hofman}}, \ and\
  \bibinfo {author} {\bibfnamefont {S.~F.}\ \bibnamefont {Lokhande}},\
  }\href@noop {} {\bibfield  {journal} {\bibinfo  {journal} {Journal of High
  Energy Physics}\ }\textbf {\bibinfo {volume} {2019}},\ \bibinfo {pages} {58}
  (\bibinfo {year} {2019})}\BibitemShut {NoStop}%
\bibitem [{\citenamefont {Martyn}\ and\ \citenamefont
  {Swingle}(2019)}]{martyn2019product}%
  \BibitemOpen
  \bibfield  {author} {\bibinfo {author} {\bibfnamefont {J.}~\bibnamefont
  {Martyn}}\ and\ \bibinfo {author} {\bibfnamefont {B.}~\bibnamefont
  {Swingle}},\ }\href@noop {} {\bibfield  {journal} {\bibinfo  {journal}
  {Physical Review A}\ }\textbf {\bibinfo {volume} {100}},\ \bibinfo {pages}
  {032107} (\bibinfo {year} {2019})}\BibitemShut {NoStop}%
\bibitem [{\citenamefont {Farhi}\ \emph {et~al.}(2014)\citenamefont {Farhi},
  \citenamefont {Goldstone},\ and\ \citenamefont {Gutmann}}]{farhi2014quantum}%
  \BibitemOpen
  \bibfield  {author} {\bibinfo {author} {\bibfnamefont {E.}~\bibnamefont
  {Farhi}}, \bibinfo {author} {\bibfnamefont {J.}~\bibnamefont {Goldstone}}, \
  and\ \bibinfo {author} {\bibfnamefont {S.}~\bibnamefont {Gutmann}},\
  }\href@noop {} {\bibfield  {journal} {\bibinfo  {journal} {arXiv preprint
  arXiv:1411.4028}\ } (\bibinfo {year} {2014})}\BibitemShut {NoStop}%
\bibitem [{\citenamefont {Grossmann}\ and\ \citenamefont
  {Rosenhauer}(1967)}]{grossmann_temperature_1967}%
  \BibitemOpen
  \bibfield  {author} {\bibinfo {author} {\bibfnamefont {S.}~\bibnamefont
  {Grossmann}}\ and\ \bibinfo {author} {\bibfnamefont {W.}~\bibnamefont
  {Rosenhauer}},\ }\href {\doibase 10.1007/BF01326224} {\bibfield  {journal}
  {\bibinfo  {journal} {Zeitschrift fur Physik}\ }\textbf {\bibinfo {volume}
  {207}},\ \bibinfo {pages} {138} (\bibinfo {year} {1967})}\BibitemShut
  {NoStop}%
\bibitem [{\citenamefont {Aleksandrowicz}\ \emph {et~al.}(2019)\citenamefont
  {Aleksandrowicz}, \citenamefont {Alexander}, \citenamefont {Barkoutsos},
  \citenamefont {Bello}, \citenamefont {Ben-Haim}, \citenamefont {Bucher},
  \citenamefont {Cabrera-Hern{\'a}ndez}, \citenamefont {Carballo-Franquis},
  \citenamefont {Chen}, \citenamefont {Chen}, \citenamefont {Chow},
  \citenamefont {C{\'o}rcoles-Gonzales}, \citenamefont {Cross}, \citenamefont
  {Cross}, \citenamefont {Cruz-Benito}, \citenamefont {Culver}, \citenamefont
  {Gonz{\'a}lez}, \citenamefont {Torre}, \citenamefont {Ding}, \citenamefont
  {Dumitrescu}, \citenamefont {Duran}, \citenamefont {Eendebak}, \citenamefont
  {Everitt}, \citenamefont {Sertage}, \citenamefont {Frisch}, \citenamefont
  {Fuhrer}, \citenamefont {Gambetta}, \citenamefont {Gago}, \citenamefont
  {Gomez-Mosquera}, \citenamefont {Greenberg}, \citenamefont {Hamamura},
  \citenamefont {Havlicek}, \citenamefont {Hellmers}, \citenamefont {Herok},
  \citenamefont {Horii}, \citenamefont {Hu}, \citenamefont {Imamichi},
  \citenamefont {Itoko}, \citenamefont {Javadi-Abhari}, \citenamefont
  {Kanazawa}, \citenamefont {Karazeev}, \citenamefont {Krsulich}, \citenamefont
  {Liu}, \citenamefont {Luh}, \citenamefont {Maeng}, \citenamefont {Marques},
  \citenamefont {Mart{\'\i}n-Fern{\'a}ndez}, \citenamefont {McClure},
  \citenamefont {McKay}, \citenamefont {Meesala}, \citenamefont {Mezzacapo},
  \citenamefont {Moll}, \citenamefont {Rodr{\'\i}guez}, \citenamefont
  {Nannicini}, \citenamefont {Nation}, \citenamefont {Ollitrault},
  \citenamefont {O'Riordan}, \citenamefont {Paik}, \citenamefont {P{\'e}rez},
  \citenamefont {Phan}, \citenamefont {Pistoia}, \citenamefont {Prutyanov},
  \citenamefont {Reuter}, \citenamefont {Rice}, \citenamefont {Davila},
  \citenamefont {Rudy}, \citenamefont {Ryu}, \citenamefont {Sathaye},
  \citenamefont {Schnabel}, \citenamefont {Schoute}, \citenamefont {Setia},
  \citenamefont {Shi}, \citenamefont {Silva}, \citenamefont {Siraichi},
  \citenamefont {Sivarajah}, \citenamefont {Smolin}, \citenamefont {Soeken},
  \citenamefont {Takahashi}, \citenamefont {Tavernelli}, \citenamefont
  {Taylor}, \citenamefont {Taylour}, \citenamefont {Trabing}, \citenamefont
  {Treinish}, \citenamefont {Turner}, \citenamefont {Vogt-Lee}, \citenamefont
  {Vuillot}, \citenamefont {Wildstrom}, \citenamefont {Wilson}, \citenamefont
  {Winston}, \citenamefont {Wood}, \citenamefont {Wood}, \citenamefont
  {W{\"o}rner}, \citenamefont {Akhalwaya},\ and\ \citenamefont
  {Zoufal}}]{Qiskit}%
  \BibitemOpen
  \bibfield  {author} {\bibinfo {author} {\bibfnamefont {G.}~\bibnamefont
  {Aleksandrowicz}}, \bibinfo {author} {\bibfnamefont {T.}~\bibnamefont
  {Alexander}}, \bibinfo {author} {\bibfnamefont {P.}~\bibnamefont
  {Barkoutsos}}, \bibinfo {author} {\bibfnamefont {L.}~\bibnamefont {Bello}},
  \bibinfo {author} {\bibfnamefont {Y.}~\bibnamefont {Ben-Haim}}, \bibinfo
  {author} {\bibfnamefont {D.}~\bibnamefont {Bucher}}, \bibinfo {author}
  {\bibfnamefont {F.~J.}\ \bibnamefont {Cabrera-Hern{\'a}ndez}}, \bibinfo
  {author} {\bibfnamefont {J.}~\bibnamefont {Carballo-Franquis}}, \bibinfo
  {author} {\bibfnamefont {A.}~\bibnamefont {Chen}}, \bibinfo {author}
  {\bibfnamefont {C.-F.}\ \bibnamefont {Chen}}, \bibinfo {author}
  {\bibfnamefont {J.~M.}\ \bibnamefont {Chow}}, \bibinfo {author}
  {\bibfnamefont {A.~D.}\ \bibnamefont {C{\'o}rcoles-Gonzales}}, \bibinfo
  {author} {\bibfnamefont {A.~J.}\ \bibnamefont {Cross}}, \bibinfo {author}
  {\bibfnamefont {A.}~\bibnamefont {Cross}}, \bibinfo {author} {\bibfnamefont
  {J.}~\bibnamefont {Cruz-Benito}}, \bibinfo {author} {\bibfnamefont
  {C.}~\bibnamefont {Culver}}, \bibinfo {author} {\bibfnamefont {S.~D. L.~P.}\
  \bibnamefont {Gonz{\'a}lez}}, \bibinfo {author} {\bibfnamefont {E.~D.~L.}\
  \bibnamefont {Torre}}, \bibinfo {author} {\bibfnamefont {D.}~\bibnamefont
  {Ding}}, \bibinfo {author} {\bibfnamefont {E.}~\bibnamefont {Dumitrescu}},
  \bibinfo {author} {\bibfnamefont {I.}~\bibnamefont {Duran}}, \bibinfo
  {author} {\bibfnamefont {P.}~\bibnamefont {Eendebak}}, \bibinfo {author}
  {\bibfnamefont {M.}~\bibnamefont {Everitt}}, \bibinfo {author} {\bibfnamefont
  {I.~F.}\ \bibnamefont {Sertage}}, \bibinfo {author} {\bibfnamefont
  {A.}~\bibnamefont {Frisch}}, \bibinfo {author} {\bibfnamefont
  {A.}~\bibnamefont {Fuhrer}}, \bibinfo {author} {\bibfnamefont
  {J.}~\bibnamefont {Gambetta}}, \bibinfo {author} {\bibfnamefont {B.~G.}\
  \bibnamefont {Gago}}, \bibinfo {author} {\bibfnamefont {J.}~\bibnamefont
  {Gomez-Mosquera}}, \bibinfo {author} {\bibfnamefont {D.}~\bibnamefont
  {Greenberg}}, \bibinfo {author} {\bibfnamefont {I.}~\bibnamefont {Hamamura}},
  \bibinfo {author} {\bibfnamefont {V.}~\bibnamefont {Havlicek}}, \bibinfo
  {author} {\bibfnamefont {J.}~\bibnamefont {Hellmers}}, \bibinfo {author}
  {\bibfnamefont {{\L}.}~\bibnamefont {Herok}}, \bibinfo {author}
  {\bibfnamefont {H.}~\bibnamefont {Horii}}, \bibinfo {author} {\bibfnamefont
  {S.}~\bibnamefont {Hu}}, \bibinfo {author} {\bibfnamefont {T.}~\bibnamefont
  {Imamichi}}, \bibinfo {author} {\bibfnamefont {T.}~\bibnamefont {Itoko}},
  \bibinfo {author} {\bibfnamefont {A.}~\bibnamefont {Javadi-Abhari}}, \bibinfo
  {author} {\bibfnamefont {N.}~\bibnamefont {Kanazawa}}, \bibinfo {author}
  {\bibfnamefont {A.}~\bibnamefont {Karazeev}}, \bibinfo {author}
  {\bibfnamefont {K.}~\bibnamefont {Krsulich}}, \bibinfo {author}
  {\bibfnamefont {P.}~\bibnamefont {Liu}}, \bibinfo {author} {\bibfnamefont
  {Y.}~\bibnamefont {Luh}}, \bibinfo {author} {\bibfnamefont {Y.}~\bibnamefont
  {Maeng}}, \bibinfo {author} {\bibfnamefont {M.}~\bibnamefont {Marques}},
  \bibinfo {author} {\bibfnamefont {F.~J.}\ \bibnamefont
  {Mart{\'\i}n-Fern{\'a}ndez}}, \bibinfo {author} {\bibfnamefont {D.~T.}\
  \bibnamefont {McClure}}, \bibinfo {author} {\bibfnamefont {D.}~\bibnamefont
  {McKay}}, \bibinfo {author} {\bibfnamefont {S.}~\bibnamefont {Meesala}},
  \bibinfo {author} {\bibfnamefont {A.}~\bibnamefont {Mezzacapo}}, \bibinfo
  {author} {\bibfnamefont {N.}~\bibnamefont {Moll}}, \bibinfo {author}
  {\bibfnamefont {D.~M.}\ \bibnamefont {Rodr{\'\i}guez}}, \bibinfo {author}
  {\bibfnamefont {G.}~\bibnamefont {Nannicini}}, \bibinfo {author}
  {\bibfnamefont {P.}~\bibnamefont {Nation}}, \bibinfo {author} {\bibfnamefont
  {P.}~\bibnamefont {Ollitrault}}, \bibinfo {author} {\bibfnamefont {L.~J.}\
  \bibnamefont {O'Riordan}}, \bibinfo {author} {\bibfnamefont {H.}~\bibnamefont
  {Paik}}, \bibinfo {author} {\bibfnamefont {J.}~\bibnamefont {P{\'e}rez}},
  \bibinfo {author} {\bibfnamefont {A.}~\bibnamefont {Phan}}, \bibinfo {author}
  {\bibfnamefont {M.}~\bibnamefont {Pistoia}}, \bibinfo {author} {\bibfnamefont
  {V.}~\bibnamefont {Prutyanov}}, \bibinfo {author} {\bibfnamefont
  {M.}~\bibnamefont {Reuter}}, \bibinfo {author} {\bibfnamefont
  {J.}~\bibnamefont {Rice}}, \bibinfo {author} {\bibfnamefont {A.~R.}\
  \bibnamefont {Davila}}, \bibinfo {author} {\bibfnamefont {R.~H.~P.}\
  \bibnamefont {Rudy}}, \bibinfo {author} {\bibfnamefont {M.}~\bibnamefont
  {Ryu}}, \bibinfo {author} {\bibfnamefont {N.}~\bibnamefont {Sathaye}},
  \bibinfo {author} {\bibfnamefont {C.}~\bibnamefont {Schnabel}}, \bibinfo
  {author} {\bibfnamefont {E.}~\bibnamefont {Schoute}}, \bibinfo {author}
  {\bibfnamefont {K.}~\bibnamefont {Setia}}, \bibinfo {author} {\bibfnamefont
  {Y.}~\bibnamefont {Shi}}, \bibinfo {author} {\bibfnamefont {A.}~\bibnamefont
  {Silva}}, \bibinfo {author} {\bibfnamefont {Y.}~\bibnamefont {Siraichi}},
  \bibinfo {author} {\bibfnamefont {S.}~\bibnamefont {Sivarajah}}, \bibinfo
  {author} {\bibfnamefont {J.~A.}\ \bibnamefont {Smolin}}, \bibinfo {author}
  {\bibfnamefont {M.}~\bibnamefont {Soeken}}, \bibinfo {author} {\bibfnamefont
  {H.}~\bibnamefont {Takahashi}}, \bibinfo {author} {\bibfnamefont
  {I.}~\bibnamefont {Tavernelli}}, \bibinfo {author} {\bibfnamefont
  {C.}~\bibnamefont {Taylor}}, \bibinfo {author} {\bibfnamefont
  {P.}~\bibnamefont {Taylour}}, \bibinfo {author} {\bibfnamefont
  {K.}~\bibnamefont {Trabing}}, \bibinfo {author} {\bibfnamefont
  {M.}~\bibnamefont {Treinish}}, \bibinfo {author} {\bibfnamefont
  {W.}~\bibnamefont {Turner}}, \bibinfo {author} {\bibfnamefont
  {D.}~\bibnamefont {Vogt-Lee}}, \bibinfo {author} {\bibfnamefont
  {C.}~\bibnamefont {Vuillot}}, \bibinfo {author} {\bibfnamefont {J.~A.}\
  \bibnamefont {Wildstrom}}, \bibinfo {author} {\bibfnamefont {J.}~\bibnamefont
  {Wilson}}, \bibinfo {author} {\bibfnamefont {E.}~\bibnamefont {Winston}},
  \bibinfo {author} {\bibfnamefont {C.}~\bibnamefont {Wood}}, \bibinfo {author}
  {\bibfnamefont {S.}~\bibnamefont {Wood}}, \bibinfo {author} {\bibfnamefont
  {S.}~\bibnamefont {W{\"o}rner}}, \bibinfo {author} {\bibfnamefont {I.~Y.}\
  \bibnamefont {Akhalwaya}}, \ and\ \bibinfo {author} {\bibfnamefont
  {C.}~\bibnamefont {Zoufal}},\ }\href {\doibase 10.5281/zenodo.2562110}
  {\enquote {\bibinfo {title} {Qiskit: An open-source framework for quantum
  computing},}\ } (\bibinfo {year} {2019})\BibitemShut {NoStop}%
\bibitem [{\citenamefont {Debnath}\ \emph {et~al.}(2016)\citenamefont
  {Debnath}, \citenamefont {Linke}, \citenamefont {Figgatt}, \citenamefont
  {Landsman}, \citenamefont {Wright},\ and\ \citenamefont
  {Monroe}}]{debnath2016demonstration}%
  \BibitemOpen
  \bibfield  {author} {\bibinfo {author} {\bibfnamefont {S.}~\bibnamefont
  {Debnath}}, \bibinfo {author} {\bibfnamefont {N.~M.}\ \bibnamefont {Linke}},
  \bibinfo {author} {\bibfnamefont {C.}~\bibnamefont {Figgatt}}, \bibinfo
  {author} {\bibfnamefont {K.~A.}\ \bibnamefont {Landsman}}, \bibinfo {author}
  {\bibfnamefont {K.}~\bibnamefont {Wright}}, \ and\ \bibinfo {author}
  {\bibfnamefont {C.}~\bibnamefont {Monroe}},\ }\href@noop {} {\bibfield
  {journal} {\bibinfo  {journal} {Nature}\ }\textbf {\bibinfo {volume} {536}},\
  \bibinfo {pages} {63} (\bibinfo {year} {2016})}\BibitemShut {NoStop}%
\bibitem [{\citenamefont {Olmschenk}\ \emph {et~al.}(2007)\citenamefont
  {Olmschenk}, \citenamefont {Younge}, \citenamefont {Moehring}, \citenamefont
  {Matsukevich}, \citenamefont {Maunz},\ and\ \citenamefont
  {Monroe}}]{Olmschenk07}%
  \BibitemOpen
  \bibfield  {author} {\bibinfo {author} {\bibfnamefont {S.}~\bibnamefont
  {Olmschenk}}, \bibinfo {author} {\bibfnamefont {K.~C.}\ \bibnamefont
  {Younge}}, \bibinfo {author} {\bibfnamefont {D.~L.}\ \bibnamefont
  {Moehring}}, \bibinfo {author} {\bibfnamefont {D.~N.}\ \bibnamefont
  {Matsukevich}}, \bibinfo {author} {\bibfnamefont {P.}~\bibnamefont {Maunz}},
  \ and\ \bibinfo {author} {\bibfnamefont {C.}~\bibnamefont {Monroe}},\ }\href
  {\doibase 10.1103/PhysRevA.76.052314} {\bibfield  {journal} {\bibinfo
  {journal} {Phys. Rev. A}\ }\textbf {\bibinfo {volume} {76}},\ \bibinfo
  {pages} {052314} (\bibinfo {year} {2007})}\BibitemShut {NoStop}%
\bibitem [{\citenamefont {M\o{}lmer}\ and\ \citenamefont
  {S\o{}rensen}(1999)}]{Molmer99}%
  \BibitemOpen
  \bibfield  {author} {\bibinfo {author} {\bibfnamefont {K.}~\bibnamefont
  {M\o{}lmer}}\ and\ \bibinfo {author} {\bibfnamefont {A.}~\bibnamefont
  {S\o{}rensen}},\ }\href {\doibase 10.1103/PhysRevLett.82.1835} {\bibfield
  {journal} {\bibinfo  {journal} {Phys. Rev. Lett.}\ }\textbf {\bibinfo
  {volume} {82}},\ \bibinfo {pages} {1835} (\bibinfo {year}
  {1999})}\BibitemShut {NoStop}%
\bibitem [{\citenamefont {Solano}\ \emph {et~al.}(1999)\citenamefont {Solano},
  \citenamefont {de~Matos~Filho},\ and\ \citenamefont {Zagury}}]{Solano99}%
  \BibitemOpen
  \bibfield  {author} {\bibinfo {author} {\bibfnamefont {E.}~\bibnamefont
  {Solano}}, \bibinfo {author} {\bibfnamefont {R.~L.}\ \bibnamefont
  {de~Matos~Filho}}, \ and\ \bibinfo {author} {\bibfnamefont {N.}~\bibnamefont
  {Zagury}},\ }\href {\doibase 10.1103/PhysRevA.59.R2539} {\bibfield  {journal}
  {\bibinfo  {journal} {Phys. Rev. A}\ }\textbf {\bibinfo {volume} {59}},\
  \bibinfo {pages} {R2539} (\bibinfo {year} {1999})}\BibitemShut {NoStop}%
\bibitem [{\citenamefont {Choi}\ \emph {et~al.}(2014)\citenamefont {Choi},
  \citenamefont {Debnath}, \citenamefont {Manning}, \citenamefont {Figgatt},
  \citenamefont {Gong}, \citenamefont {Duan},\ and\ \citenamefont
  {Monroe}}]{choi2014optimal}%
  \BibitemOpen
  \bibfield  {author} {\bibinfo {author} {\bibfnamefont {T.}~\bibnamefont
  {Choi}}, \bibinfo {author} {\bibfnamefont {S.}~\bibnamefont {Debnath}},
  \bibinfo {author} {\bibfnamefont {T.~A.}\ \bibnamefont {Manning}}, \bibinfo
  {author} {\bibfnamefont {C.}~\bibnamefont {Figgatt}}, \bibinfo {author}
  {\bibfnamefont {Z.-X.}\ \bibnamefont {Gong}}, \bibinfo {author}
  {\bibfnamefont {L.-M.}\ \bibnamefont {Duan}}, \ and\ \bibinfo {author}
  {\bibfnamefont {C.}~\bibnamefont {Monroe}},\ }\href {\doibase
  10.1103/PhysRevLett.112.190502} {\bibfield  {journal} {\bibinfo  {journal}
  {Phys. Rev. Lett.}\ }\textbf {\bibinfo {volume} {112}},\ \bibinfo {pages}
  {190502} (\bibinfo {year} {2014})}\BibitemShut {NoStop}%
\end{thebibliography}%
\renewcommand\thefigure{S\arabic{figure}}    
\setcounter{figure}{0}

\pagebreak
\onecolumngrid
\appendix

\section{Detailed formalism for Lee Yang Zeros}
We consider an N site spin Hamiltonian 
$\mathcal{H}_s$ with
an external field term 
$\mathcal{H}_B = h\sum_i \sigma^z_{i}$. 
Defining a variable $\zt=\exp(2 \beta h)$, the partition function for $N$ spins is written in terms of $\zt$

\begin{align}
    \mathcal{Z}(\beta,\mathcal{H}_s,h) &= \Tr[\exp(-\beta (\mathcal{H}_s + \mathcal{H}_B))] \nonumber\\
   &= \exp(-\beta N h)\sum_{k=0}^N p_k \zt^k 
    \label{eq:polynomial_expression}
\end{align}
where 
$p_k=\Tr_{\sum_i \langle\sigma^z_i\rangle=N-2k}\exp(-\beta\mathcal{H}_s)$ is the partition function in a zero magnetic field when $k$ spins are in the $\ket{\downarrow}$ state. In order to get this expansion we have used the commutativity of $\mathcal{H}_s$ and $\mathcal{H}_B$.
The partition function is expressed as an $N^{th}$ order polynomial 
in terms of the variable $\zt$\cite{lee_statistical_1952}, which using the fundamental theorem of algebra we can rewrite in terms of its N zeros ($\zt_j$) as\cite{yang_statistical_1952} 
\begin{align}
    \mathcal{Z}(\beta ,\mathcal{H}_s,h) = \exp(-\beta N h) p_N \ \Pi_{j=1} ^N \left(\zt-\zt_j\right),
    \label{eq:constructZ_appendix}
\end{align}
The coefficients of the polynomial are all positive numbers, thus its zeros cannot lie on the positive real axis (where the physical partition
function exists), but must instead lie in the complex plane of $h$.  Yet, if we can find the zeros, we may reconstruct the partition
function from them.
Now lets see how we can find these zeros experimentally. 

Returning to the definition of the partition function, with the magnetic field as complex quantity 
$h=h_r+i h_i$:
\begin{align}
    \mathcal{Z}(\beta ,\mathcal{H}_0,h_i) &= \Tr \exp\left(- \beta \mathcal{H}_0 -i \beta  h_i \sum_{i=1}^{n}\sigma_i^z \right)
    \label{eq:z_with_im_h}
\end{align}
where $\mathcal{H}_0 = \mathcal{H}_s + Re(\mathcal{H}_B)$.
At this point, the imaginary part resembles a
time evolution 
by a Hamiltonian $\sum_{i=1}^{n}\sigma_i^z$ with the identification $\lambda t = \beta h_i$.

A measurable quantity $L(t)$, proportional to the complex partition function $\mathcal{Z}$ can be found with the identification $\lambda t = \beta h_i$ as described in the main text. 
\begin{align}
    L(t) = \frac{1}{\mathcal{Z}_0}\Tr \exp{(-\beta {\mathcal{H}}_0 -i \lambda t \sum_{i=1}^{n}\sigma_i^z )},
\end{align}
where $ \mathcal{Z}_{0} $ is the partition function $ \Tr e^{-\beta \mathcal{H}_0}$.
In order to achieve this, a probe or an ancilla qubit is attached to the system with the coupling Hamiltonian\cite{wei_phase_2014},
\begin{align}
      \mathcal{H}' = \frac{\lambda}{2} \left(\sigma^z_{\text{probe}}  \otimes \sum_{i=1}^{n}\sigma_i^z \right).
      \label{eq:TE_Hamiltonian}
\end{align}
The ancilla is initialised to be in the $\ket{+}$ state and the system in the thermal state, here $\ket{+} = \frac{1}{\sqrt{2}} [\ket{0} +\ket{1}] $. Thus the initial density matrix of the total system is 
\begin{align}
    \rho(0) =(\ket{+}\bra{+}) \otimes 
    \frac{e^{-\beta \mathcal{H}_0}}{\mathcal{Z}_{0}}.
    \label{eq:rho0tot_appendix}
\end{align}
The time-evolved density matrix under the coupling Hamiltonian is,
\begin{align}\label{eq:rho_t}
    \rho(t) = e^{-i \mathcal{H}' t} \rho(0) e^{i \mathcal{H}' t}.
\end{align}
Since $\mathcal{H}_0 $ commutes with $\mathcal{H}_B$, the density matrix becomes
\begin{align}
    \rho(t)&=\frac{1}{2Z_{0}}\left(\ket{\uparrow}\bra{\uparrow}e^{-\beta \mathcal{H}_0}+\ket{\downarrow}\bra{\downarrow}e^{-\beta \mathcal{H}_0}\right)\nonumber\\
   &+\frac{1}{2Z_{0}}\left(\ket{\uparrow}\bra{\downarrow}e^{-\beta \mathcal{H}_0}e^{- i \lambda t \sum_{i=1}^{N}\sigma_i^z} +h.c.\right)
\end{align}
Now $L(t)$ can be extracted from the off diagonal terms of the reduced density matrix of the ancilla. From the probe spin's perspective, the off-diagonal element of its density
matrix (after tracing out the system) becomes
\begin{align}
    \rho_{\uparrow\downarrow}^{ancilla}(t) = \frac{1}{2Z_0}\Tr_{\mathrm{sys}} \exp\left(-\beta \mathcal{H}_0 - i \lambda t \sum_{i=1}^{N}\sigma_i^z\right) = \frac{1}{2}L(t)
\end{align}
Thus the real part of $L(t)$ can be extracted from the expectation value of $\sigma_z$ of the ancilla after applying Hadamard gate and similarly imaginary part of $L(t)$ can be extracted after applying $R_x(- \pi /2)$) gate. Note that for this procedure, $\mathcal{H}_0$ needs to commute with $\mathcal{H}_B$. Otherwise, we need a different coupling Hamiltonian and the implementation becomes difficult\cite{wei_phase_2014}.

The above described method can be summarised for the quantum simulation into the following three steps:

\begin{enumerate}
    \item Prepare the system,including the probe in its initial state according to Eq.~\ref{eq:rho0tot_appendix}.
    \item Time-evolve with the Hamiltonian Eq.~\ref{eq:TE_Hamiltonian}, where the time evolution operator is $U(t) = \exp{(-i\mathcal{H}^{'} t)}$.
    \item Measure the off-diagonal components of the ancilla density matrix to get $L(t)$. Zeros of $L(t)$ are the zeros of the partition function $\mathcal{Z}$
\end{enumerate}


\subsection{Lee-Yang Zeros for the Ising model}

\begin{figure}[ht]
    \includegraphics[clip=true,trim=00 60 20 40,width=0.9\textwidth]{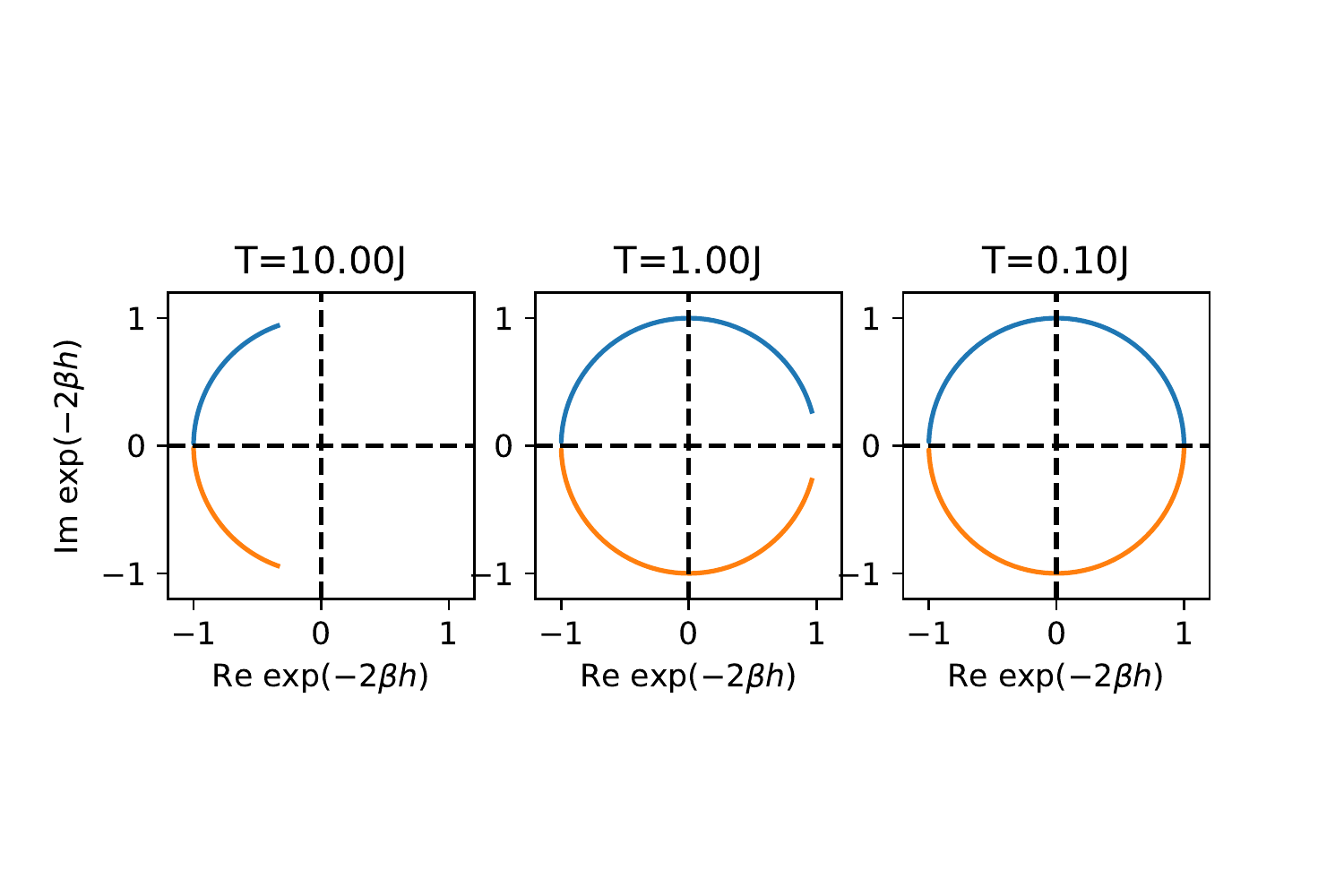}
    \caption{Lee-Yang zeros for the classical Ising model.}
    \label{fig:zeros_classical_ising}
\end{figure}

The one dimensional Ising Hamiltonian with periodic boundary condition for N sites is 
\begin{align}
\mathcal{H}=-J\sum_{i=1}^{N} \sigma^z_{i}\sigma^z_{i+1}-h\sum_{i=1}^{N} \sigma^z_{i}. 
\end{align}
For the ferromagnetic case, where $J>0$, the Lee-Yang zeros are purely imaginary in $h$ and are given in terms of $\zt=\exp(-2 \beta h)$ as


\begin{align}
    \zt& = -e^{-4\beta J}\left(1+\cos(k_n)\right) + \cos(k_n)
     \pm i\sqrt{\left(1-e^{-4\beta J}\right)\left[\sin(k_n)^2+e^{-4\beta J}(1+\cos(k_n))^2\right]}
    \label{ising_anlytc}
\end{align}
for $k_n=\frac{\pi(2n-1)}{N}$, where $N$ is the number of sites. This result may be obtained from a transfer matrix formalism. Since the zeros are purely imaginary in $h$, $\zt=\exp(-2\beta h)$ lies on the unit circle, as shown in Fig.~\ref{fig:zeros_classical_ising}. As the temperature is increased, the distribution of zeros collapses to a point where $2\beta h=\pi$. At lower temperatures, the zeros complete the circle, pinching the real axis at the critical temperature, corresponding to $\zt_{\rm crit}=e^{-2\beta_{\rm crit}h}$.

\subsection{Lee-Yang Zeros for the XY model}
We next consider the XY model, where the
Hamiltonian is
\begin{align}
H = J\sum_i( \sigma^x_i \sigma^x_{i+1} +\sigma^y_i \sigma^y_{i+1} ) + h\sum_{i=1}^{N} \sigma^z_{i}.
\end{align}
This model can be diagonalised through a Jordan-Wigner transformation followed by Fourier transformation. The Lee-Yang zeros in h are properly complex, with the imaginary part given by $\cos(2 \beta h_i)=-1$ and the real part given by $h_r = -2 J \cos(k)$, where $k$ are the quasi-momenta used in the Fourier-basis representation of the chain \cite{tong_lee-yang_2006-1}.

\subsection{Lee-Yang Zeros for the two site XXZ model}
The two site XXZ Hamiltonian is
\begin{equation}
    H = J(\sigma^x_1 \sigma^x_2 +\sigma^y_1 \sigma^y_2 ) + J_z(\sigma^z_1 \sigma^z_2) + h( \sigma^z_1 + \sigma^z_2).
\end{equation}{}
Here, the Lee-Yang zeros occur at the values of $h$ shown in Table~\ref{tab:2sitexxz_lyz}. We note that 
if $h_1 = h_r +ih_i$ corresponds to a zero, then so does $-h_1$. Thus if $\zt_1 = \exp(2 \beta h_1 )$ is one solution to the polynomial, then the other solution is given by $\zt_2 = \exp(-2 \beta h_1 )$.
We may also read off from the partition function that the constraints that Ising-type zeros are found when $\cosh(2\beta J) < \exp(-2\beta J_z)$, and vice versa for XY-type zeros.

\begin{table}[h]
\begin{tabular}{ |c|c|c| } 
 \hline
  Type & $h_r/h_i$ & $h_r/h_i$ \\ 
\hline
Ising & $h_r=0$ & $ \cos(2\beta h_i) = - \cosh(2\beta J) \exp(2\beta J_z)$ \\ 
XY & $ 2\beta h_i=(2n +1) \pi $ & $\cosh(2\beta h_r)= \cosh(2\beta J) \exp(2\beta J_z)$ \\ 
 \hline
\end{tabular}
\caption{Real/imaginary parts of the complex magnetic field $h$ where the zeros of the 2-site XXZ model occur.}
\label{tab:2sitexxz_lyz}
\end{table}

\section{Circuit for the preparation of the TFD state of the 2-site XXZ model}
\label{appendix:TFD_XXZ}

\begin{figure*}[htpb]
\centering
\hspace{0.in}
\vspace{0.1in}
\Qcircuit @C=0.7em @R0.8em {
& \gate{H} &\ctrl{2}& \qw&\multigate{1}{XX(\theta_1)}&\multigate{1}{YY(\theta_1)}&\gate{Z(\theta_2)}&\qw&\qw&\qw&\multigate{1}{ZZ(\theta_3)}&\multigate{1}{XX(\theta_4)}&\qw&\qw&\qw \\
&\gate{H} &\qw &\ctrl{2} &\ghost{XY(\theta_1)} &\ghost{YY(t_1)}&\gate{Z(\theta_2)}&\qw  & \qswap &\qw &\ghost{ZZ(\theta_3)} &\ghost{XX(\theta_4)} &\qswap &\qw &\qw\\
 &\qw &\gate{X}&\qw&\multigate{1}{XX(\theta_1)}&\multigate{1}{YY(\theta_1)}&\gate{Z(\theta_2)} &\qw &\qswap  \qwx &\qw &\multigate{1}{ZZ(\theta_3)}&\multigate{1}{XX(\theta_4)}  \qw&\qswap \qwx &\qw&\qw\\
&\qw &\qw  & \gate{X}&\ghost{XY(\theta_1)} &\ghost{YY(\theta_1)}&\gate{Z(\theta_2)}&\qw&\qw&\qw &\ghost{ZZ(\theta_3)} &\ghost{XX(\theta_4)}&\qw &\qw&\qw
\gategroup{1}{2}{4}{4}{.7em}{--}
\gategroup{1}{10}{4}{14}{.7em}{--} 
}
\vspace{0.1in}
\hspace{0.in}
\Qcircuit @C=0.7em @R0.8em {
&\multigate{1}{XX(\theta_5)}&\multigate{1}{YY(\theta_5)}&\gate{Z(\theta_6)}&\qw&\qw&\qw&\multigate{1}{ZZ(\theta_7)}&\multigate{1}{XX(\theta_8)}&\qw&\qw&\qw&\qw \\
 &\ghost{XY(\theta_5)} &\ghost{YY(\theta_5)}&\gate{Z(\theta_6)}&\qw &\qswap  &\qw&\ghost{ZZ(\theta_7)} &\ghost{XX(\theta_8)}&\qw  &\qswap \qw &\qw&\qw \\
&\multigate{1}{XX(\theta_5)}&\multigate{1}{YY(\theta_5)}&\gate{Z(\theta_6)}&\qw&\qswap \qwx &\qw&\multigate{1}{ZZ(\theta_7)}&\multigate{1}{XX(\theta_7)} &\qw &\qswap \qwx &\qw\qw&\qw \\
&\ghost{XY(\theta_5)} &\ghost{YY(\theta_5)}&\gate{Z(\theta_6)}&\qw&\qw&\qw&\ghost{ZZ(\theta_7)} &\ghost{XX(\theta_8)}  &\qw&\qw  &\qw
\gategroup{1}{2}{4}{4}{.7em}{--}
\gategroup{1}{8}{4}{12}{.7em}{--}
}

\caption{Circuit to prepare a thermofield double state of the 2 site XXZ model with 8 parameters $\theta_1, \ldots \theta_8$. Here $XX(\theta)=\exp{(-i\theta \sigma_x \sigma_x)}, YY(\theta)= \exp{(-i\theta\sigma_y \sigma_y)}, ZZ(\theta)= \exp{(-i\theta\sigma_z \sigma_z)}, Z(\theta)= \exp{(-i \frac{\theta}{2} \sigma_z)}$}
\label{fig:TFD_XXZ}
\end{figure*}
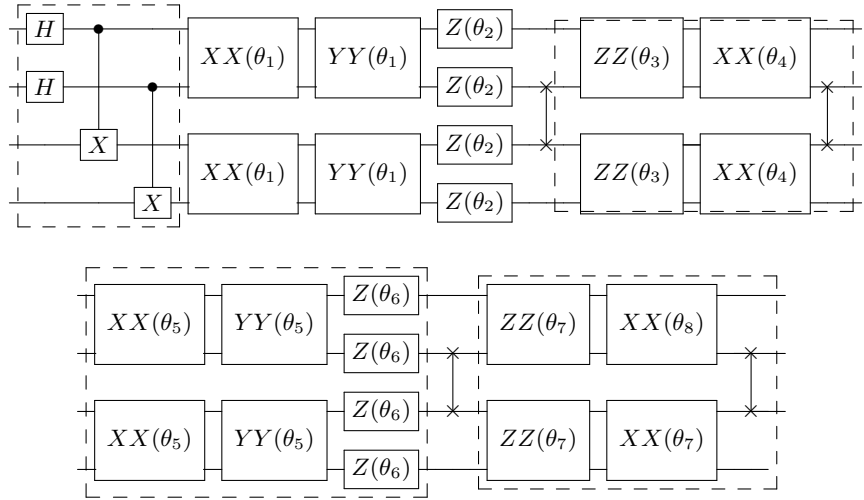
\begin{figure*}
    \includegraphics[width=0.99\textwidth]{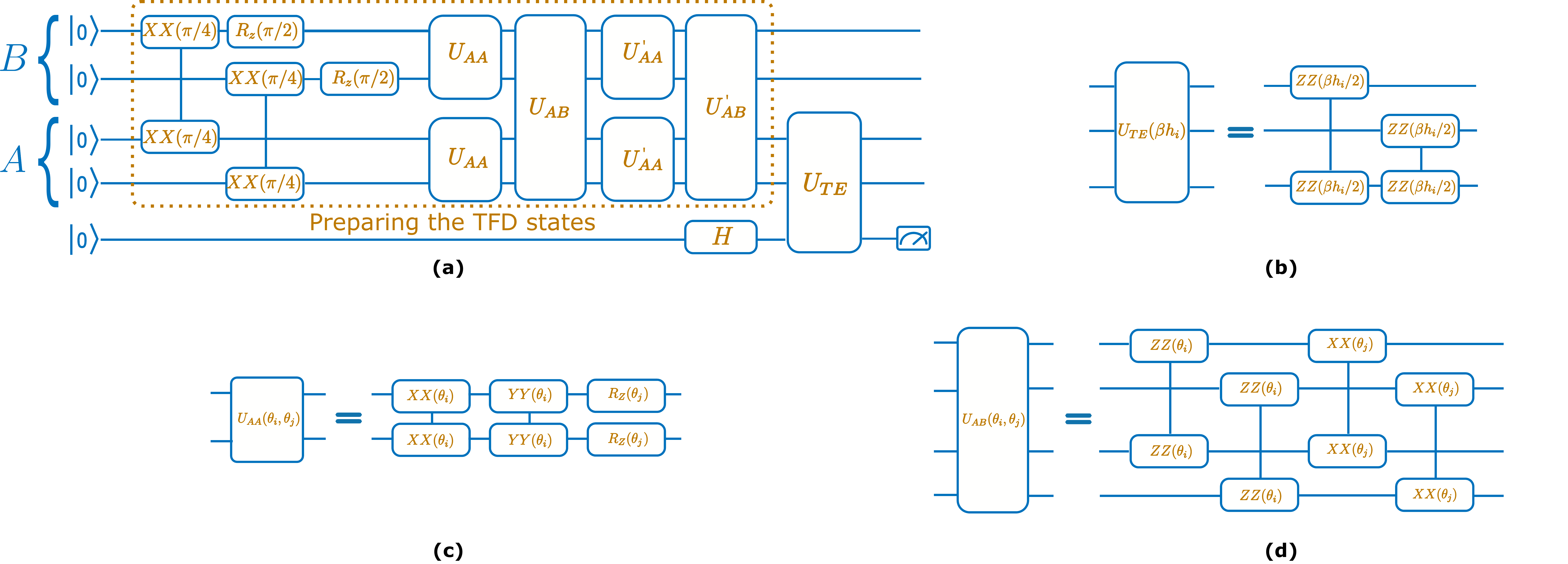}
	\caption{Experimental realization of the full circuit. 
	(a) Overview indicating the operative parts: TFD preparation via an alternating operator ansatz, followed by time evolution under the interaction Hamiltonian.
	(b)-(d) Decomposition of the time evolution and variational ansatz unitary operations into hardware native gates.}
	\label{fig:experimental_circuit}
\end{figure*}

For finding the location of the Lee Yang zeros of a system with Hamiltonian $H_A$, we need to prepare a thermal density matrix at a particular temperature, $\rho_{\beta}=\mathcal{Z}^{-1} e^{-\beta H_{A}}$. 
This can be obtained by tracing out one subsystem from a TFD state which is defined in an expanded Hilbert space that contains two copies of the system:
\begin{align}
|\operatorname{TFD}(\beta)\rangle=\frac{1}{\sqrt{Z}} \sum_{n} e^{-\beta E_{n} / 2}|n\rangle_{A}|n\rangle_{B},
\end{align}
where $H_{A}|n\rangle_{A}=E_{n}|n\rangle_{A}$. To prepare this TFD, a variational ansatz involving alternating time evolution under Hamiltonians $H_{A}^{'}$, $H_{B}^{'}$, and $H_{AB}$ is constructed \cite{wu_variational_2019,zhu_variational_2019} as shown in Fig. \ref{fig:TFD_XXZ}. $H_{A}^{'}$ and $H_{B}^{'}$ act identically on systems A and B respectively and are close in structure to the Hamiltonian $H_A$. $H_{AB}$ is an inter-system Hamiltonian which entangles both systems. At the start, the system is prepared in a maximally entangled
Bell state $\frac{1}{\sqrt{2}}(\ket{00}+\ket{11})$, which is the ground state of $ H_{AB}$.  This ansatz starts from the correct thermal state at infinite temperature and if enough time evolution layers are applied, could produce the ground state at zero temperature. In our case, $H_A^{'}$ was modified from $H_A$ with an additional parameter in order to shorten the circuit depth. 

The optimisation for the parameters was performed classically; the resulting circuit parameters $\theta_1$ to $\theta_8$  are given in the Table~\ref{tab:t1-t8}.
Fig.~\ref{fig:experimental_circuit} shows the hardware implementation of the final circuit.
\begin{table}[htpb]
    \centering
\begin{tabular}{ |c|c|c|c|c|c|c|c|c| } 
 \hline
  J  &$-\theta_1$& $-\theta_2/2$&$-\theta_3$ &$-\theta_4$ &$-\theta_5$ & $-\theta_6/2$ &$-\theta_7$ & $-\theta_8$ \\
 \hline
0.9 & 0.409 & 0.785 & 0.480 &1.660 &0.395 & 0.785 &0.739 &1.178 \\ 
0.96 & 1.178&  0.392 &  0.555 &  1.427 &  1.092 &  0.392 &  0.694 & -0.360 \\
1.03 & 0.993& 0.785& 1.014&  1.210&  1.060&  0.785&  0.933&  0.392 \\
1.06 & 0.922 & 1.486 &  0.438 &  0.887 &  0.678 &  1.446 &  0.624 &  1.165 \\
1.15 & 0.958&  0.948& 1.008&  1.133&  0.752&  0.753& 0.590& 1.187 \\
1.20 & 0.972&  0.968&  0.990&  1.163&  0.772& 0.772& 0.589&  1.182\\

 \hline
\end{tabular}
\caption{$\theta_1$ to $\theta_8$ parameters for two site XXZ TFD preparation.}
    \label{tab:t1-t8}
\end{table}

\section{Simulation of noise in experiment}
\begin{figure*}[htpb]
    \centering
    \includegraphics[width=1 \textwidth]{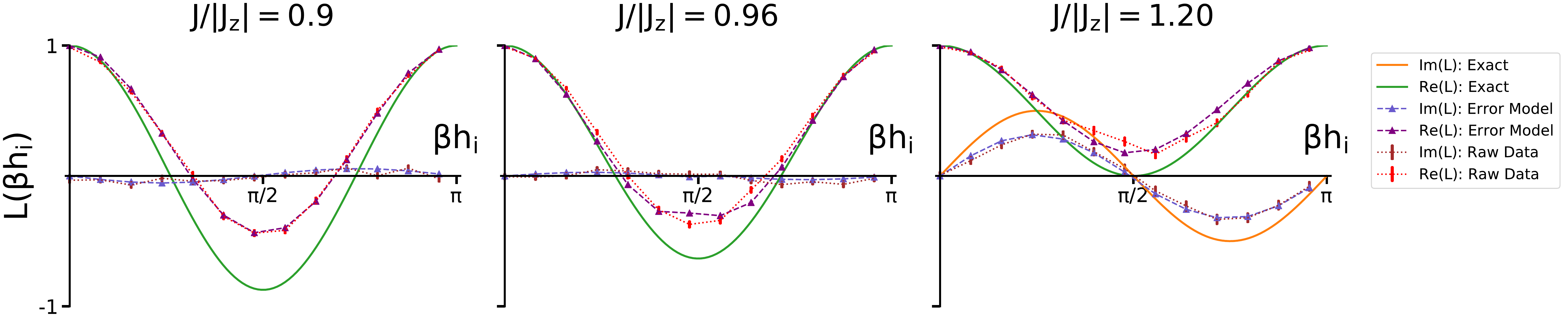}
    \caption{ L($h_i$) obtained from the linear shift error model is compared with the actual data and exact values. }
    \label{fig:linear shift error model}
\end{figure*}

On the ion trap quantum computer, the native two qubit gate used to implement entangling operations is ideally defined as $XX_{i,j}(t) = \exp{(-i t \sigma_i^x\sigma_j^x)}$. In practice, the physical operation deviates from the ideal unitary. Previously the effect of random under or over rotations in the XX gate on the preparation of TFD states has been explored \cite{zhu_variational_2019}. Here we study the effect of systematic shifts in the angle using a 'linear shift' error model which assumes a modified XX gate with two parameters $a$ and $b$:
\begin{align}
    XX_{i,j}(t) \rightarrow \left[Z_i(bt_{trim}) \otimes Z_j(bt_{trim})\right] \cdot XX_{i,j}(at_{trim})\cdot(\sigma^x \otimes \sigma^x)^n  ,
\end{align}
where $Z(\theta)=\exp{(-i (\theta/2) \sigma^z)}$. Here $t_{trim}$ is the angle trimmed from $t$ by adding or substracting $\pi/2$ '$n$' times so that the trimmed angle is in the range (-$\pi/4$,$\pi/4)$. This is in accordance with how the XX gates are implemented on an ion-trap quantum computer. We performed a simulation in which all the XX gates in the circuit were replaced by this modified gate\cite{Qiskit}. The optimised value of $a$ and $b$ are obtained by minimising least square distance to data points and are given in Table~\ref{tab: linear shift error}. The corresponding plots are shown in Fig.~\ref{fig:linear shift error model}. We see that the error is well modeled by the simulation.

\begin{table}[htpb]
    \centering
\begin{tabular}{ |c|c|c| } 
 \hline
 $J$  &$a$& $b$ \\ 
 \hline
 0.9 & 0.99121641 & -0.47829858\\
 0.96 & 1.12953451 & 0.042765\\
 1.20 & 0.99011104 & -0.36491532\\
\hline
\end{tabular}
\caption{Optimised values of the parameters for linear shift error model. 
}
    \label{tab: linear shift error}
\end{table}

\section{Post-Selecting Experimental Data}
Two post-selection schemes can be applied at the end of the circuit for calculating Lee Yang Zeros after preparing the TFD state corresponding to the XXZ model. 


Method 1: For the XXZ model, $H = J\sum_i(\sigma^x_i \sigma^x_{i+1} +\sigma^y_i \sigma^y_{i+1} ) + J_z\sum_i(\sigma^z_i \sigma^z_{i+1}) + h \sum_i \sigma^z_i$, $\sum_i \sigma^z_i$ is a good quantum number. Therefore, in the TFD state for this model, 
\begin{equation}
    \ket{\Psi}=\frac{1}{Z_{\beta}}\sum_j e^{-\beta E_j/2}\ket{\phi_j}_A\ket{\phi_j}_B,
\end{equation}

we have $\sum_i \sigma^z_{i,A}=\sum_i \sigma^z_{i,B}$. Any runs of the circuit resulting in measurements that do not satisfy this condition can be discarded.

Method 2: Measuring the real part of the LYZ curves involves putting the ancilla qubit in the $(\ket{0}+\ket{1})/\sqrt{2}$ state, followed by the operation $\exp\big(-i\frac{\theta}{2}\sigma^z_a\sum_i \sigma^z_i\big)$, where $\sigma^z_a$ acts on the ancilla qubit and the sum is over the qubits in subsystem A of the TFD state, and an $R_y(-\pi/2)$ on the ancilla before measurement. For the 2-site XXZ model, this can be decomposed as:

\begin{figure}[h]
\mbox{
\Qcircuit @C=1em @R=1.7em {
\lstick{\ket{0}} & \gate{H}   &\ctrl{1}    &\ctrl{2}   &\gate{R_y(-\pi/2)}   & \meter\\
&\ustick{a_1}    \qw     &\gate{Rz(-2\theta)}   &\qw            &\gate{Rz(\theta)} & \qw & \meter\\
&\ustick{a_2}    \qw     &\qw          &\gate{Rz(-2\theta)}   &\gate{Rz(\theta)} & \qw & \meter
}
}
\end{figure}

Now, if $\sum_i \sigma^z_{i,A}=0 $, that is, $\sigma^z_{a_1}\neq\sigma^z_{a_2}$, the controlled rotations will cancel each other resulting in no phase generated on the ancilla qubit. After the final $R_y(-\pi/2)$, the ancilla qubit will then return to $0$ with probability 1. Therefore, any runs of the circuit that result in a measurement in which the ancilla qubit is 1 but $\sigma^z_{a_1}\neq\sigma^z_{a_2}$ should be discarded.



\begin{figure*}[htpb]
    \centering
    \includegraphics[width=1 \textwidth]{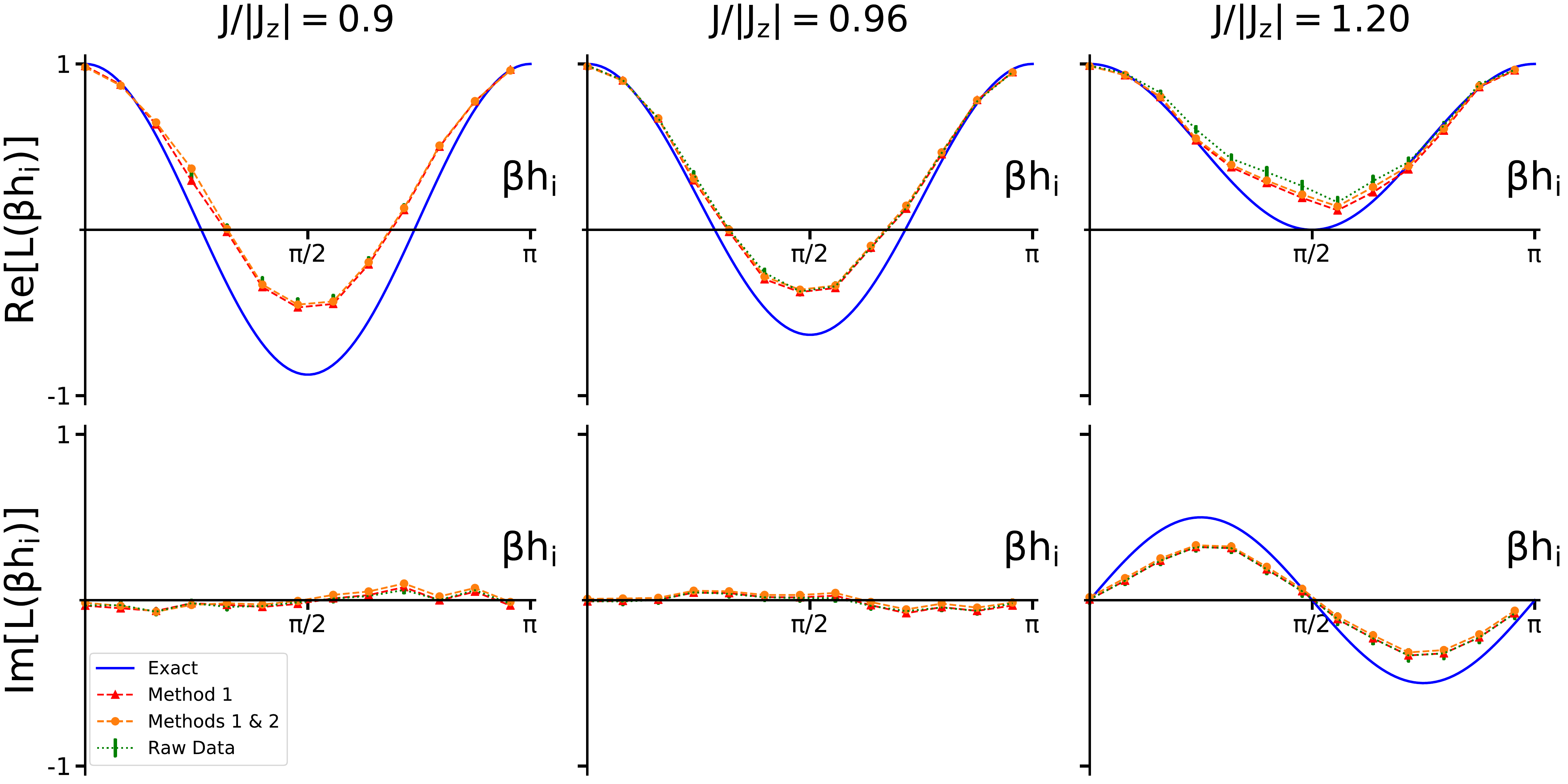}
    \caption{Comparing different methods of data post selection shows that they only slightly improve the accuracy of the experimental data. }
    \label{fig:post-selection data}
\end{figure*}

Fig. \ref{fig:post-selection data} shows the effect of the two post-selection schemes on the measured data.

\section{Fisher Zeros}
Following the study of Lee Yang zeros, Fisher looked into the partition function zeros in terms of complex temperature \cite{Fisher1965lectures}. We refer to Fisher zeros as zeros in terms of inverse temperature $\beta$. For the XXZ model, analytical expressions are difficult to find. But for the two limiting cases, Ising and XY model analytical expressions are known. For Ising model with Hamiltonian,
\begin{align}
H = J \sum_i^N \sigma^z_i \sigma^z_{i+1},
\end{align}
Fisher zeros are found at \cite{jones1966complex}
\begin{align}
\beta = -\frac{1}{4J}\ln{\tan^2[{\frac{\pi}{N}(k + 1/2) }]} \pm i \frac{\pi}{4J}(2m+1)
\end{align}
where $k = \{0,1,...,N-1\}$ and $m$ is any integer.
For the XY model diagonalisation is performed after mapping into a fermionic space using a Jordan-Wigner transformation. In order to compare Fisher zeros to the analytical expression we have considered the XY model with the boundary term which does not have any Jordan Wigner string. 

\begin{align}
H = J \sum_i^{N-1} (\sigma^x_i \sigma^x_{i+1}  + \sigma^y_i \sigma^y_{i+1}) + H_b
\end{align}
$H_b$ is the boundary term $H_b=2J[\sigma_+^1 \sigma_z ....\sigma_z \sigma_-^N + \sigma_-^1 \sigma_z ....\sigma_z \sigma_+^N]$, where $\sigma_{\pm} = (\sigma_x \pm i \sigma_y)/2 $. This gives the fermionic energies to be $4J\cos(\frac{2\pi}{N} k)$ where $ k = \{0,1,...,N-1\} $. Hence equating partition function to zero, Fisher zeros are found to be on the imaginary axis at 
\begin{align}
\beta = -i \frac{(2m+1)\pi}{4J\cos(\frac{2\pi}{N} k)}
\end{align}
where $m$ is an integer.



\end{document}